\documentclass[structabstract]{aa} 

\usepackage{graphicx,rotating,epsfig}
\usepackage{txfonts,times,latexsym,amssymb}
\usepackage{natbib}
\usepackage{longtable,lscape,multirow,supertabular}
\usepackage{color}
\usepackage{mathtools}

\begin{document}

\title{On the properties of dust and gas in the environs of V838 Monocerotis\thanks{{\it Herschel} is an ESA space observatory with science instruments provided by European-led Principal Investigator consortia and with important participation from NASA.}} 
%\subtitle{}essentially

\author{
  K. M. Exter\inst{1,2,3}
    \and
  N.L.J. Cox\inst{1,4}
  \and
  B. M. Swinyard\inst{5,6}
  \and
  M. Matsuura\inst{5,7}
  \and
  A. Mayer\inst{8}
  \and
  E. De Beck\inst{9,10}
   \and
  L. Decin\inst{1,11}
  }

\institute{
  Institute of Astronomy, KU Leuven, Celestijnenlaan 200D, BUS 2401, 3001 Leuven, email katrina.exter@kuleuven.be 
  \and
  Herschel Science Centre, European Space Astronomy Centre, ESA, P.O.Box 78, Villanueva de la Ca\~nada, Spain
  \and
  ISDEFE, Beatriz de Bobadilla 3, 28040 Madrid, Spain
  \and
  present address: Universit\'e de Toulouse, UPS-OMP, IRAP, 31028, Toulouse, France; CNRS, IRAP, 9 Av. colonel Roche, BP 44346, F-31028 Toulouse, France
  \and
  Department of Physics and Astronomy, University College London, Gower Street, London WC1E 6BT, UK
  \and
  RAL Space, Rutherford Appleton Laboratory, Chilton, Didcot, Oxfordshire, OX11 0QX, UK
  \and
  School of Physics and Astronomy, Cardiff University, The Parade, Cardiff CF24 3AA, UK
  \and
  Department of Astrophysics, University of Vienna, T{\"u}rkenschanzstrasse 17, 1180 Vienna, Austria
  \and
  Department of Earth and Space Sciences, Chalmers University of Technology, Onsala Space Observatory, SE-439 92 Onsala, Sweden
  \and
  Max-Planck Institut f\"ur Radioastronomie, Auf dem H\"ugel 69, 53121 Bonn, Germany
  \and
  Astronomical Institute Anton Pannekoek, Universiteit van Amsterdam, Science Park 904, NL-1090GE, Amsterdam, The Netherlands
  }

\date{Received; accepted}

\abstract{% context
}{% aims
We aim to probe the close and distant circumstellar environments of the stellar outburst object V838\,Mon. 
}{% methods
\emph{Herschel} far-infrared imaging and spectroscopy were taken at several epochs to probe the central point source and the extended
environment of V838\,Mon. PACS and SPIRE maps were used to obtain photometry of the dust immediately around  V838\,Mon, and in the surrounding infrared-bright region. These maps were fitted in 1d and 2d to measure the temperature, mass, and $\beta$ of the two dust sources. PACS and SPIRE spectra were used to detect emission lines from the extended atmosphere of the star, which were then modelled to study the physical conditions in the emitting material.  HIFI spectra were taken to measure the kinematics of the extended atmosphere but unfortunately yielded no detections. 
}{% results
Fitting of the far-infrared imaging of V838\,Mon reveals 0.5--0.6~M$_\odot$ of  $\approx19$~K dust in the environs  ($\approx2.7$\,pc) surrounding V838\,Mon. The surface-integrated infrared flux (signifying the thermal light echo), and derived dust properties do not vary significantly between the different epochs. We measured the photometry of the point source. As the peak of the SED (Spectral Energy Distribution) 
 lies outside the {\it Herschel} spectral range, it is only by incorporating data from other observatories and previous epochs that we can usefully fit the SED; with this we explicitly assume no evolution of the point source between the epochs. We find that warm dust with a temperature $\sim300$\,K distributed over a radius of 150--200\,AU.\\ 
We fit the far-infrared lines of CO arising from the point source, from an extended environment around V838\,Mon. Assuming a model of a spherical shell for this gas, we find that the CO appears to arise from two temperature zones: a cold zone (T$_{kin}\approx18$\,K) that could be associated with the ISM or possibly with a cold layer in the outermost part of the shell, and a warm (T$_{kin}\approx400$\,K) zone that is associated with the extended  environment of V838\,Mon within a region of radius of  $\approx210$\,AU. The SiO lines arise from a warm/hot zone.  We did not fit the lines of H$_2$O as they are far more dependent on the model assumed. 
}{% conclusion (opt)
}

\keywords{infrared: stars -- novae, cataclysmic variables -- stars: individual (V838 Monocerotis) -- ISM: clouds -- ISM: dust}

\maketitle

\section{Introduction}

The object V838 Monocerotis (hereafter V838\,Mon) is one of the most enigmatic observed in stellar astrophysics in recent decades. It came to attention when it underwent a powerful eruptive outburst in 2002 (\citealt{2012PASA...29..466R}),  increasing in luminosity by a factor of 100 over a period of three months. Immediately following this event a spectacular light echo was formed from the outburst light reflecting off the surrounding dust (\citealt{2003Natur.422..405B}). Post outburst, V838\,Mon varied through spectral types F, G, and K, to a very cool M-giant (\citealt{2003MNRAS.343.1054E}). \citet{Loebman:2015el} found the star to show features of mixed spectral type, from M3 dwarf to mid-L supergiant: the atmosphere is obviously not a conformist.  

V838\,Mon is situated in a molecular cloud and shares its neighbourhood with a sparse cluster of young B stars (e.g.\,\citealt{AB07}). 
It has a B-type companion, located 28--250\,AU from V838\,Mon (Munari et al., 2007; Tylenda et al., 2009) which was engulfed by the material expanding out from V838\,Mon after its outburst (Bond, 2006; Munari et al., 2007; Kolka et al., 2009). 
Photometry from before the outburst will be a mix of the companion star and the precursor; and using pre-outburst photometry to fix the status of the precursor requires a reliable measure of the distance. 
Analysis of high-resolution {\it HST} polarimetry images of the light-echo led \cite{2003Natur.422..405B} to place a lower limit of 6\,kpc to the distance of V838 Mon. Working on the same HST material, \cite{Tylenda04} revised the distance to $8\pm2$ kpc. \cite{2005A&A...434.1107M} favoured an even larger distance of closer to 10\,kpc. 
However, the latest value is $6.2\pm0.6$\,kpc, measured by \cite{2008AJ....135..605S} on the basis of a detailed study of the light echo. 
At this distance, the companion completely dominates the pre-outburst photometry (\citealt{AB07}), and hence the precursor must have been the fainter star.  We will adopt a value of 6.2\,kpc in this paper. 

The type of star V838 Mon was before the outburst, and the nature of the outburst itself, is as yet unresolved. It is made difficult by the sparsity of data pre-outburst, and the uncertain photometry (i.e.\ the presence of the secondary star). It has been suggested that the star was massive and hot (e.g.\ \citealt{2005A&A...434.1107M}) and the outburst was a thermonuclear event; that the outburst was a born-again event in a solar-mass white dwarf in close orbit with a lower-mass companion (\citealt{Lawlor05}); that it was pre-/main-sequence star of 5--10\,M\sun\ (\citealt{2005A&A...441.1099T}); and that it was a main-sequence star which suffered a merger with planets or a lower-mass star (e.g.\ \citealt{retteretal07}, \citealt{2003ApJ...582L.105S}). \citet{2006A&A...451..223T} discuss the theories to date to explain V838\,Mon, and similar outbursts in two other stars, M31 RV  (\citealt{
1989ApJ...341L..51R}) and V4332\,Sgr (\citealt{1999AJ....118.1034M}), and they favour the stellar merger mode; in this paper we assume this model is correct, although it makes little difference to many of our calcuations.

There have been numerous studies of V838\,Mon (the point source and the circumstellar region) in atomic lines and molecules from the optical to the mid-IR and a few in the far-IR. The spectrum has been evolving dynamically and in terms of composition from the first to the most recent observations. The response of the star to the outburst seems to have been to develop an expanding photosphere and a multi-component circumstellar shell (\citealp{2004ApJ...607..460L}, \citealp{2007A&A...467..269G},  \citealp{2009A&A...503..899T}), this material flowing out and some falling back.  The velocities are generally from some tens to 200\,km\,s$^{-1}$, and temperatures range from about 2000\,K in the photosphere, to a few 100\,K in the extended circumstellar region some 200--300\,AU out.

The dust in the molecular cloud surrounding V838\,Mon can be seen as the light echo sweeps through, and \emph{HST} images show a very chaotic region of loops and whirls and cavities (\citealp{2003Natur.422..405B}, \citealp{2005MNRAS.358.1352C}). The appearance of the surrounding dust has led to the speculation that the outburst was not an isolated event. \citet{2004A&A...427...193V} found that IRAS and MSX observations show evidence for multiple mass-loss events prior to the 2002 outburst: a 7\arcmin$\times10$\arcmin\ dust shell is visible in the IRAS data, and the MSX data also show an extended object (1.5\arcmin\ diameter) at the position of V838\,Mon. However, estimates of the combined mass of this dust and gas are in the range 10s--100 solar masses (e.g. \citealt{2006ApJ...644L..57B}; \citealt{2011A&A...529A..48K}; \citealt{2012A&A...548A..23T}). This  would rule out all of this material having an origin in V838\,Mon itself: instead, we are looking at the interstellar cloud from which V838\,Mon, and other members of its  cluster, were formed.  CO studies (\citealp{2011A&A...529A..48K}, \citealp{Kaminski:2007gg}) also show this extended material, confirming it is massive and moreover cool ($\sim20$\,K).

{\it Spitzer} observations from 2004/5  at 24\,$\mu$m, 70\,$\mu$m and 160\,$\mu$m show extended emission distributed  over an arcminute from the star, and moreover that this is coincident with the {\it HST} light echo (\citealt{2006ApJ...644L..57B}).  This infra-red ``light echo'' is believed to arise from reprocessed thermal emission from dust grains ($\sim$1~M$_\odot$) heated by outward propagating UV--visible photons. Unresolved emission at 24\,$\mu$m and 70\,$\mu$m (PSFs of 6\arcsec\ and 24\arcsec) is also seen in the {\it Spitzer} data at the position of the central star; moreover the flux has increased between 2004/5 and 2007 (\citealt{2008ApJ...683L.171W}), which is suggested to indicate the presence of hot dust freshly condensed in the outburst ejecta. Gemini images taken in 2007 in the mid-IR, at 11.2\,$\mu$m and 18.1\,$\mu$m, do not show evidence of  extended emission over radial distances of up to 15\arcsec\ from the central source (\citealt{2008ApJ...683L.171W}).

It is clear that V838\,Mon is an intriguing star, which is varying with time and which is interesting to study across most of the spectrum. We decided to take advantage of the opportunity to observe  V838\,Mon in the further reaches of the IR with the good spectral and spatial resolution that the instruments onboard \emph{Herschel} offered. We report on our imaging and spectral study of V838\,Mon and its near-by environment. 

We note that in this paper we will use the word "extended" to refer to the emission we detect arising from the molecular cloud in which V838\,Mon is found (i.e. the region which is coincident with the HST light echo, with a distribution of an arcminute or so from the star). "Point source" refers to anything detected by \emph{Herschel} as a point source: the FWHM of the PACS 70\,$\mu$m beam is  $\sim9$\arcsec (55000\,AU at a 6.2\,kpc distance), and that of the SPIRE 500\,$\mu$m beam is  $\sim37$\arcsec. These large sizes mean that the point source includes the star, its circumstellar  environment, and probably also  ISM.

\begin{table}[ht!]
\caption{The \emph{Herschel} observations.}\label{tab:obs}
\centering
\begin{tabular}{lll}
\multicolumn{3}{c}{PACS} \\ \hline
Date, epoch\tablefootmark{a}   & Obs.\ id.\tablefootmark{b} & $\lambda$ reference ($\mu$m)\tablefootmark{c}  \\ 
m.d.20xx& & \\ \hline 
05.07.11, 1 & 1342220129,30  & phot: 70, 160 \\  
              & 1342220131,2   & phot: 100, 160  \\
              & 1342220136     & spec: 79.36, 162.81 \\
              &                & 165.95, 177.77   \\
              & 1342220137     & spec: 179.53, 180.78  \\
              &                & 169.68, 71.07  \\
              & 1342220138     & spec: 118.58,108.07  \\
05.01.12, 2 & 1342245206,7   & phot: 70,160 \\
              & 1342245208,9   & phot: 100,160  \\
10.16.12, 3 & 1342253505,6   & phot: 70,160  \\
              & 1342253507,8   & phot: 100,160 \\ \hline
\multicolumn{3}{c}{SPIRE} \\\hline
09.18.10, 1 & 1342204851     & phot: 250, 350, 500 \\
04.24.11, 1 & 1342219551     & spec: 194--671 \\
04.30.12, 2 & 1342245152     & phot:  250, 350, 500 \\
10.14.12, 3 & 1342253396     & phot:  250, 350, 500    \\ \hline               
\multicolumn{3}{c}{HIFI} \\\hline
&& Setting; band\hfill$\nu$ (GHz)\tablefootmark{d}  \\ \hline
09.13.11, 1.5 & 1342228563& A; 1b  \hfill 606.752-610.884 \\
                                                &                                       &\hfill 618.745-622.877\\
09.30.11, 1.5  & 1342229897 & B; 2a   \hfill 678.314-682.445 \\
                      &                                         &       \hfill 690.307-694.439\\
09.14.11,1.5 & 1342228604 & C; 5a \hfill 1149.527-1153.658 \\
                      &                                         &\hfill 1161.520-1165.651  \\
10.09.11, 1.5 & 1342230404 & D; 7a  \hfill 1759.995-1762.558 \\
                                            &                                           &\hfill  1767.243-1769.806\\
09.29.11, 1.5 & 1342229843 & E; 7b \hfill 1789.211-1791.774 \\
                                            &                                           &\hfill  1796.459-1799.022\\ \hline
\end{tabular}
\tablefoot{
\tablefoottext{a}{Epoch: a counter based on the PACS observation dates; for SPIRE and HIFI the closest epoch to these is indicated}
\tablefoottext{b}{Obs.\ Id refers to the unique reference number given to all \emph{Herschel} observations}
\tablefoottext{c}{phot=photometer, spec=spectrometer}
\tablefoottext{d}{WBS ranges in LSB and USB}
}
\end{table}

\begin{figure*}[ht!]
  \centering
   \includegraphics[scale=0.46]{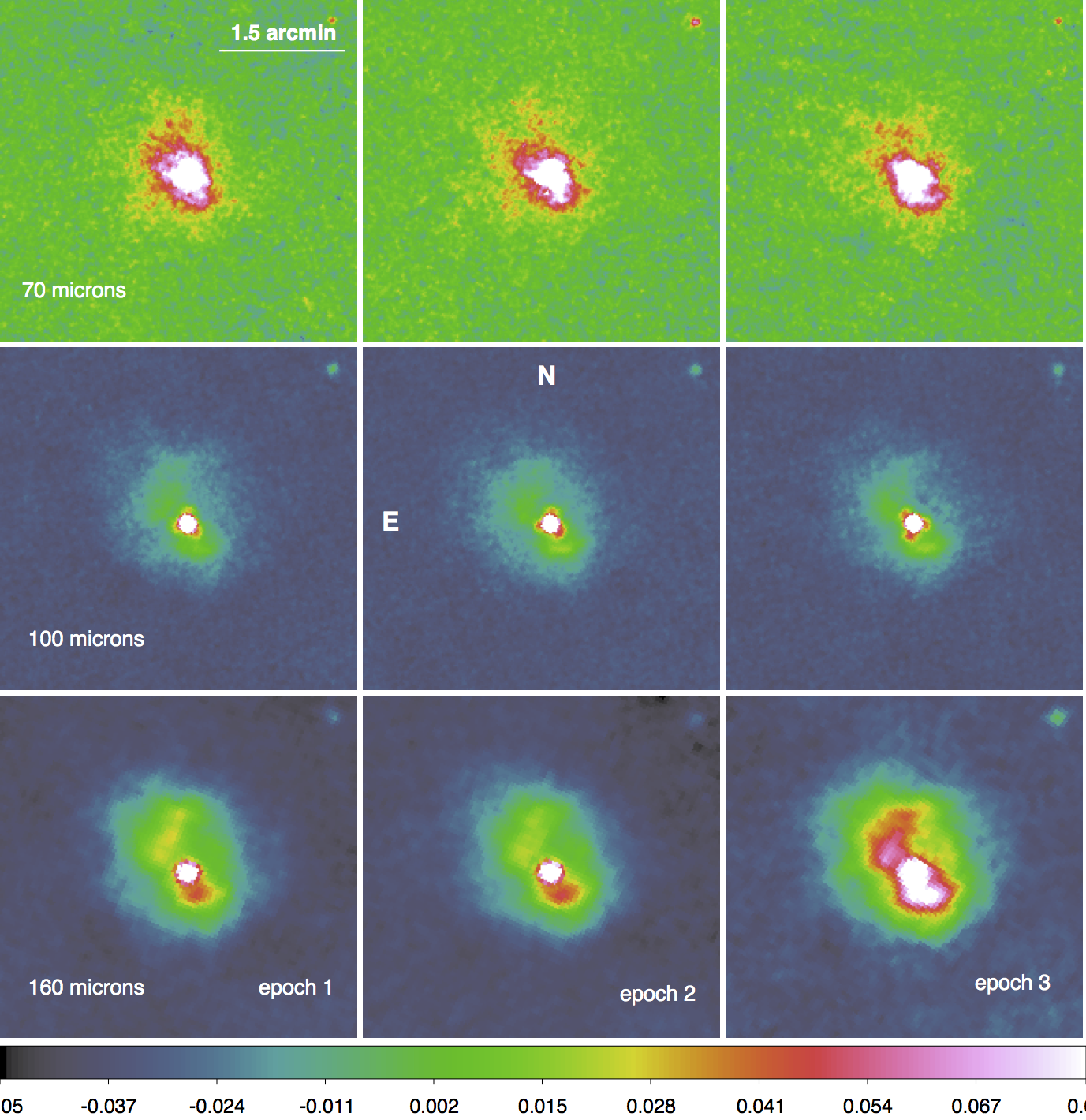}
      \includegraphics[scale=0.46]{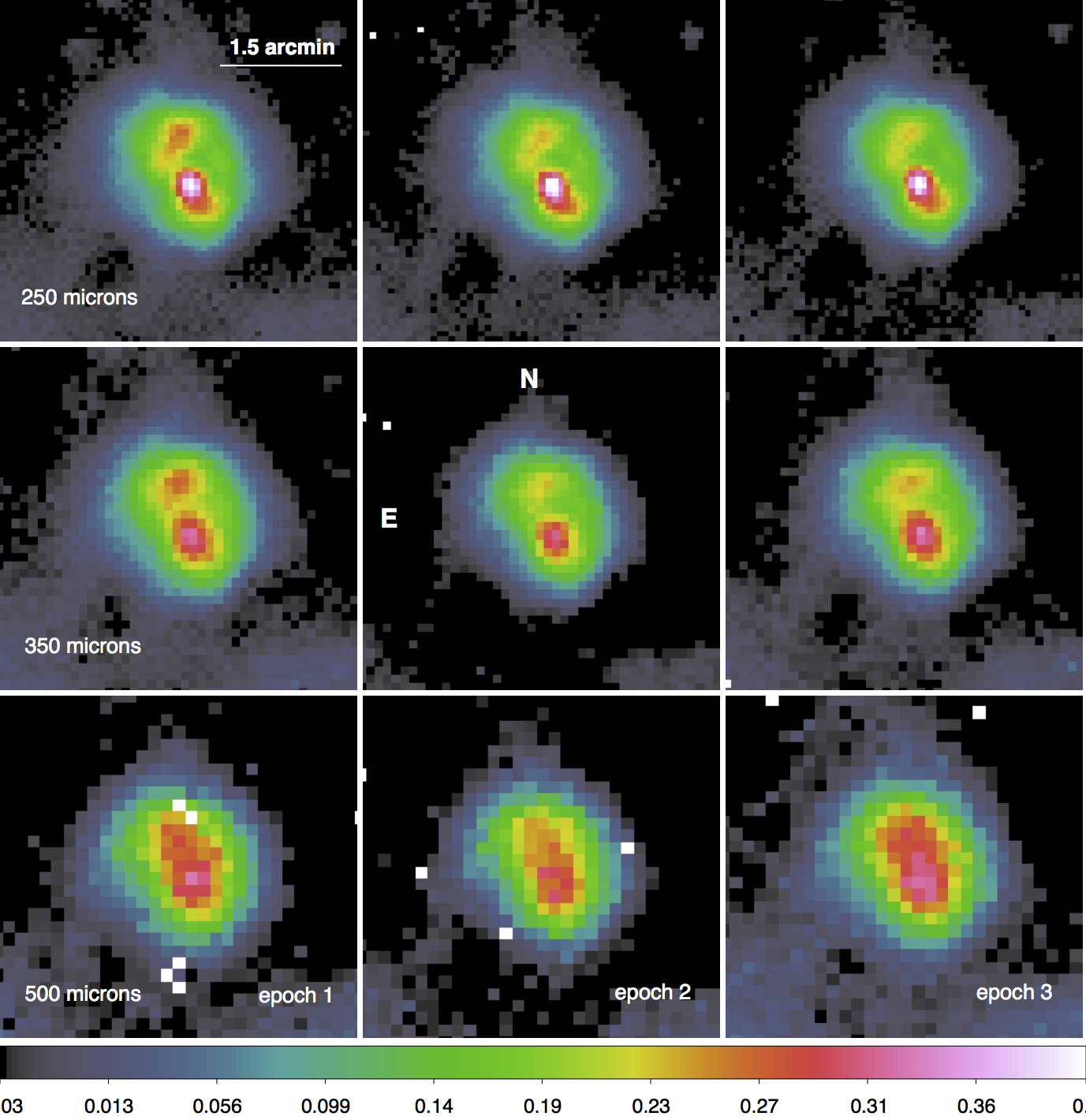}
   \caption{{\bf Top to bottom}  PACS (top three) and SPIRE (bottom three) maps at the indicated wavelengths and epochs. The flux scaling, in increasing wavelength order, is: PACS -0.05 to 0.08/0.2/0.4 Jy\,beam$^{-1}$; SPIRE -0.03 to 0.8/0.6/0.4 Jy\,beam$^{-1}$. The brightest point on each map is V838\,Mon; the tri-lobal shape of the PACS beam can be seen on the 100\,$\mu$m maps. N--E is indicated in the central maps.} 
   \label{fig:firstmaps}
\end{figure*}

\section{\emph{Herschel} observations and data processing}

Photometric imaging with the \emph{Herschel} (\citealt{herschel}) instruments PACS (\citealt{pacs}) and SPIRE (\citealt{spire}) was obtained on three epochs spread over 1.5 years, and additionally PACS, SPIRE, and HIFI (\citealt{hifi}) spectroscopy was obtained on the first epoch. Details of the observations are given in Table~\ref{tab:obs}. In this section we describe the pipeline data reduction and the production of the science-quality maps and spectra.

\subsection{PACS photometry}\label{sec:pacsphot}

For PACS we use the scan map AOT (astronomer observation template)  with two orthogonal (45\degr and 135\degr) scanning angles, four scan legs with a length of 12\arcmin.4 each and a cross-scan step of 155\arcsec, and medium scan speed.  We obtained four maps at each scan angle: 70 and 160\,$\mu$m simultaneously, and 100 and 160\,$\mu$m simultaneously. The final maps are then produced by concatenating the frames from both the scan and cross-scan observations, resulting in maps with a uniform coverage of a square area of just over 6\arcmin\ on a side. The separate PACS observations were taken back-to-back and within a few months of the SPIRE observations. 

The  data were  processed up to Level 1 with the HIPE (\citealt{hipe}) track 12 pipeline for scan maps and with calibration tree 56. The data were then transferred to be processed with Scanamorphos (\citealt{Roussel2013}, version 23). The blue and green maps were processed with the default options in the ``minimap'' mode; for the red frames we also used the option ``flat'' in Scanamorphos when combing the data from all four datasets to produce a clean background. Glitch detection was done in Scanamorphos. The pixel sizes were chosen for the best results in the subsequent data analysis: 1.0\arcsec\ at 70\,$\mu$m, 1.5\arcsec\ at 100\,$\mu$m and 3.0\arcsec\ at 160\,$\mu$m. Finally, we subtracted the mean of the background flux to set all the maps to the same zero level.  

There were slight shifts of the world coordinate system (WCS) of the maps between the epochs -- the position of V838\,Mon (and a faint background point source)  differed by a few pixels. This could be a consequence of slight telescope astrometry errors. Since we later make a direct comparison of the morphology of the extended emission between epochs, we moved the WCS of epochs 2 and 3 to that of epoch 1, independently for each band. This was achieved by fitting the position of V838\,Mon (using the HIPE task {\sl sourceFitting}) and resetting the WCS of the maps to the appropriate sky and pixel coordinates.   

The calibration uncertainties for PACS photometry are 5\%\ from the models and 2\%\ reproducibility (information provided on the PACS documentation page on the \emph{Herschel} portal), and a recent report of a comparison of different map-makers shows that those commonly used on PACS data perform to within 10\%\ of each other for extended emission (\citealt{Paladini2013}). The errors in the measured photometry for the point and extended sources are in fact dominated by the sky noise and the difficulties in disentangling the two sources. 

The PACS maps are shown in Fig.\,\ref{fig:firstmaps}. For ease of comparison between epochs and wavelengths we converted the units to Jy/beam. (For the conversion we used the beam areas taken from a PACS Photometer PSF report of 4 April 2012 (``bolopsf\_20.pdf'') from the PACS documentation page on the HSC portal. We used beam area values of 40, 54, and 145 sq.\,arcsec\ for the blue, green, and red maps.)

\subsection{SPIRE photometry}\label{sec:spirephot}

For SPIRE we used the standard small map AOT with a repetition factor three. The observations were done in three filters (PSW=250\,$\mu$m, PMW=350\,$\mu$m, PLW=500\,$\mu$m) simultaneously, and produced maps with a uniform coverage over a 5\arcmin\ diameter circle.

The SPIRE data were reduced in HIPE (track 12, calibration tree spire\_cal\_12\_2) using the pipeline recipe for small maps (with the pipeline script parameter {\sl applyExtendedEmissionGain} set to True). The  Planck offsets were applied via the task {\sl zeroPointCorrection}. We opted for map pixel sizes slightly smaller than the default values as this worked better for our subsequent analysis: 4.5\arcsec, 6.25\arcsec\ and 9.0\arcsec\ for PSW, PMW, and PLW respectively. We produced the maps calibrated for extended emission (units of MJy\,sr$^{-1}$, and used in the analysis of the extended emission) and those calibrated for point sources (units of Jy\,beam$^{-1}$, and used to measure the photometry of the point source).  We also created maps with Scanamorphos, using the standard reduction path and created maps with pixel sizes of 3\arcsec, 5\arcsec\ and 7\arcsec\ for PSW, PMW, and PLW respectively and calibrated for point sources. 
We subtracted the mean of the background flux to set all the maps to the same zero level (subsequent comparison of the maps was then easier). Background source subtraction was not necessary.  
The calibration uncertainties for SPIRE photometry are 4\%\ systematic and 1.5\%\ random  for a point source, and an additional 4\%\ for the extended emission (\citealt{2013MNRAS.433.3062B}): we adopted a value of 6\%\ overall. As with the PACS photometry, the errors are in fact dominated by the sky noise and the difficulties in disentangling the two sources. 
The SPIRE maps (those produced in HIPE and calibrated for point sources) are shown in Fig.\,\ref{fig:firstmaps}.  

\begin{figure*}[pt!]
  \centering
   \includegraphics[scale=0.75]{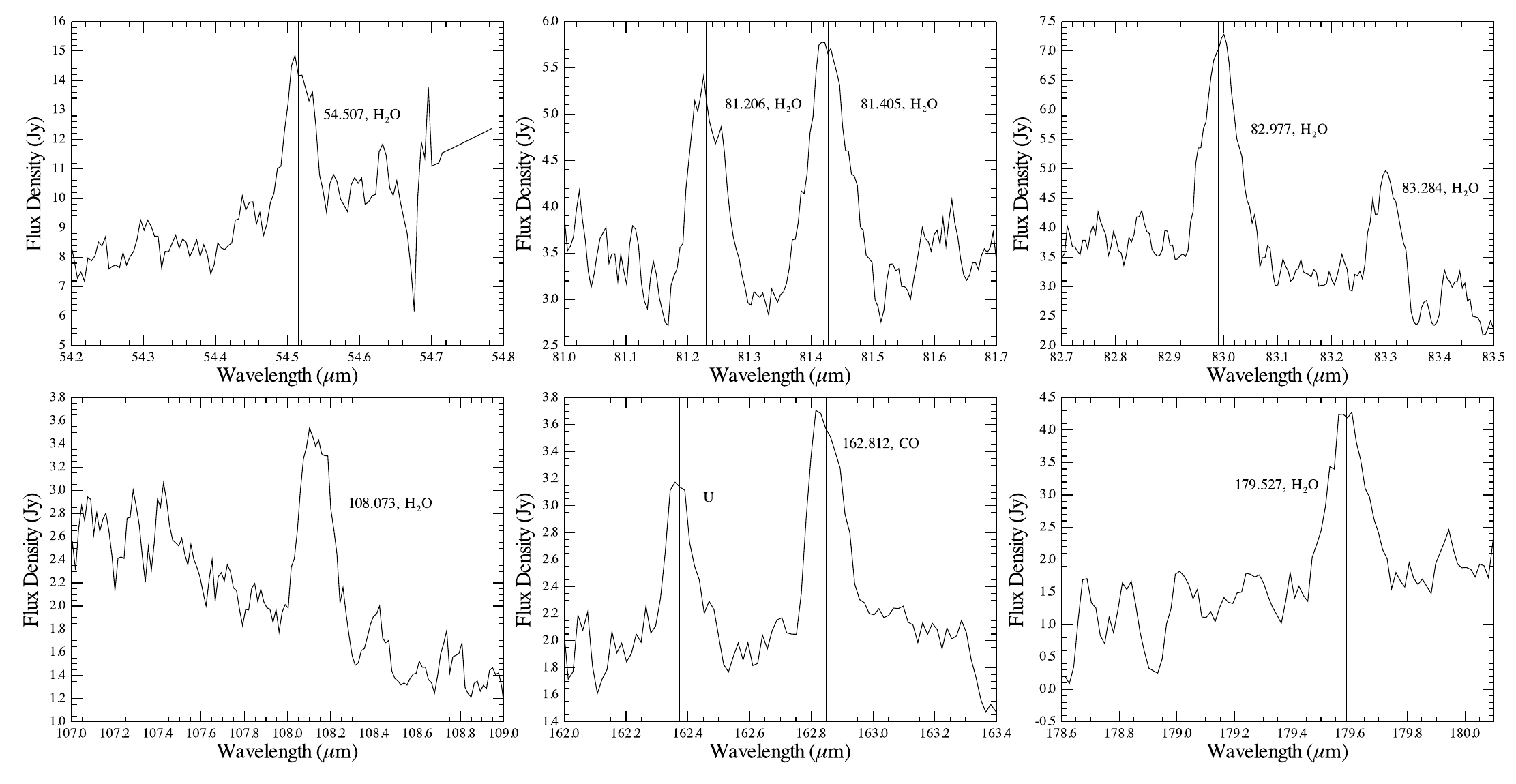}
   \caption{Examples of PACS spectra of V838\,Mon, with identifications indicated. The feature at 162.36\,$\mu$m (1846.4675~GHz) has not been unambiguously identified.  
   }
   \label{fig:pacsspecex}
\end{figure*}

\subsection{PACS spectroscopy}\label{sec:pacsspec}

We obtained ten PACS spectral line scans, split between three concatenated AOTs. We used the chop-nod (medium throw) Pointed mode, performing line-scan spectroscopy with line repetition factors between one and three and a chop-nod repetition factor~one at the (default\footnote{The instrumental resolution is  about: 0.02\,$\mu$m, 100\,km\,s$^{-1}$  (up to 60\,$\mu$m); 0.04\,$\mu$m, 120\,km\,s$^{-1}$ (up to 100\,$\mu$m); and 0.11--0.12\,$\mu$m, 190\,km\,s$^{-1}$ (up to 190\,$\mu$m)}) high-density spectral sampling setting. The reference wavelengths of the scans are given in Table\,\ref{tab:obs}. The wavelength range of the scans is about 2\,$\mu$m in the red and 1\,$\mu$m in the blue. 

Our PACS observations consist of 20 discrete segments: the ten requested line scans and ten accompanying parallel scans.  
We used the background normalisation pipeline script for line scans to reduce the data, in HIPE track 12 with calibration tree version 60. This pipeline uses the telescope background to calibrate the data, and is recommended for observations of long duration and of low continuum flux levels, this latter being the case for V838\,Mon.

Diffuse emission from the extended source around V838\,Mon pervades the entire spectroscopy field-of-view. 
The PACS integral field unit (IFU) extends by $\sim47$\arcsec$\times47$\arcsec\ (covered by a $5\times5$ array of spaxels), while our PACS maps show that the extended emission covers a patchy area of 1\arcmin--2\arcmin. This means that flux from the extended emission may be present in the spaxels containing the point source:  see App.\,\ref{sec:app3}. 
To measure the level of contamination of this diffuse emission on the stellar spectrum of V838\,Mon, we compared the flux levels of the spectra of the outermost spaxels from our on-source pointings (i.e.\ the spaxels sitting on the extended emission only, with no point-source contribution) to the spectra from the off-source/blank sky pointings (which contain only telescope background emission) to see how much flux from the extended emission could have been picked up  in the on-source observations: there is an excess of 0.1--0.4\,Jy in the outer spaxels of the on-source pointings, which we believe comes from the extended source. Under the assumption that this flux is the same over the whole field-of-view, 
we removed this contribution from the spectra in the final 20 cubes produced by the pipeline. For each cube, we averaged the spectra from the faintest ten outer spaxels;  fit a low-order polynomial to the average spectrum; and  subtracted the fit from the entire cube. The subsequently-extracted point source spectrum of V838\,Mon should then be clear of contamination from the extended emission. It is important to do this before applying the next step -- the point source calibration -- because that step requires the point source to contain only emission from the itself.

We  discovered that V838\,Mon was off-centred during our observations: it is located almost exactly between the central spaxel and one of its neighbours.  
This has consequences for the extraction of the spectrum of the star. The standard task provided in the PACS pipeline 
scripts to extract a correctly-calibrated spectrum of a point source ({\sl extractCentralSpectrum}) requires it to be 
located  close to the centre of the central spaxel. However, we were able to modify the task to work on V838\,Mon, and while the resulting spectrum will not be as well corrected as a properly-centred source, it is superior to applying no correction. For details see App.~\ref{sec:app6}. 

Examples of the PACS spectra of V838\,Mon are shown in Fig.\,\ref{fig:pacsspecex} with some spectral line identifications. 
The absolute flux calibration uncertainty (RMS, combined absolute, reproduceability, and pointing scatter) is about 15\% peak-to-peak\footnote{See the PACS Spectrometer Performance Document on the PACS documentation pages on the HSC portal.}. For a faint source such as V838\,Mon, there is an additional uncertainty of the order $\pm$0.7\,Jy on the level of the continuum in a point-source corrected spectrum\footnote{This information will be provided at a later date also on the PACS public web-pages and on the HSC Legacy Library portal.}; we note that this is not also an additional uncertainty on the integrated line fluxes. The higher uncertainty in this continuum level due to our modification of the point source correction task and the background subtraction will take this to about $\pm$1Jy.

\subsection{SPIRE spectroscopy}\label{sec:spirespec}

We obtained a single SPIRE-FTS pointing with sparse image sampling in the high spectral resolution mode and with 23 scan repetitions of the FTS mechanism to build up the signal-to-noise in the interferogram. The SPIRE FTS simultaneously covers the (SSW) short wavelength band (190--313\,$\mu$m; 31--52\,cm$^{-1}$; 957--1577\,GHz) and (SLW) long wavelength band (303--650\,$\mu$m; 15--33\,cm$^{-1}$; 461--989\,GHz), which both have an 
unapodised spectral resolution of 0.048 cm$^{-1}$. The resulting data set consists of a spectrum from the central bolometer (i.e.\ V838\,Mon) and spectra from the surrounding concentric circles of bolometers. The centre-to-centre distance of the bolometers is 33\arcsec\ in the SSW band and 51\arcsec\ in the SLW band.

\begin{figure*}
  \centering
  \includegraphics[scale=1]{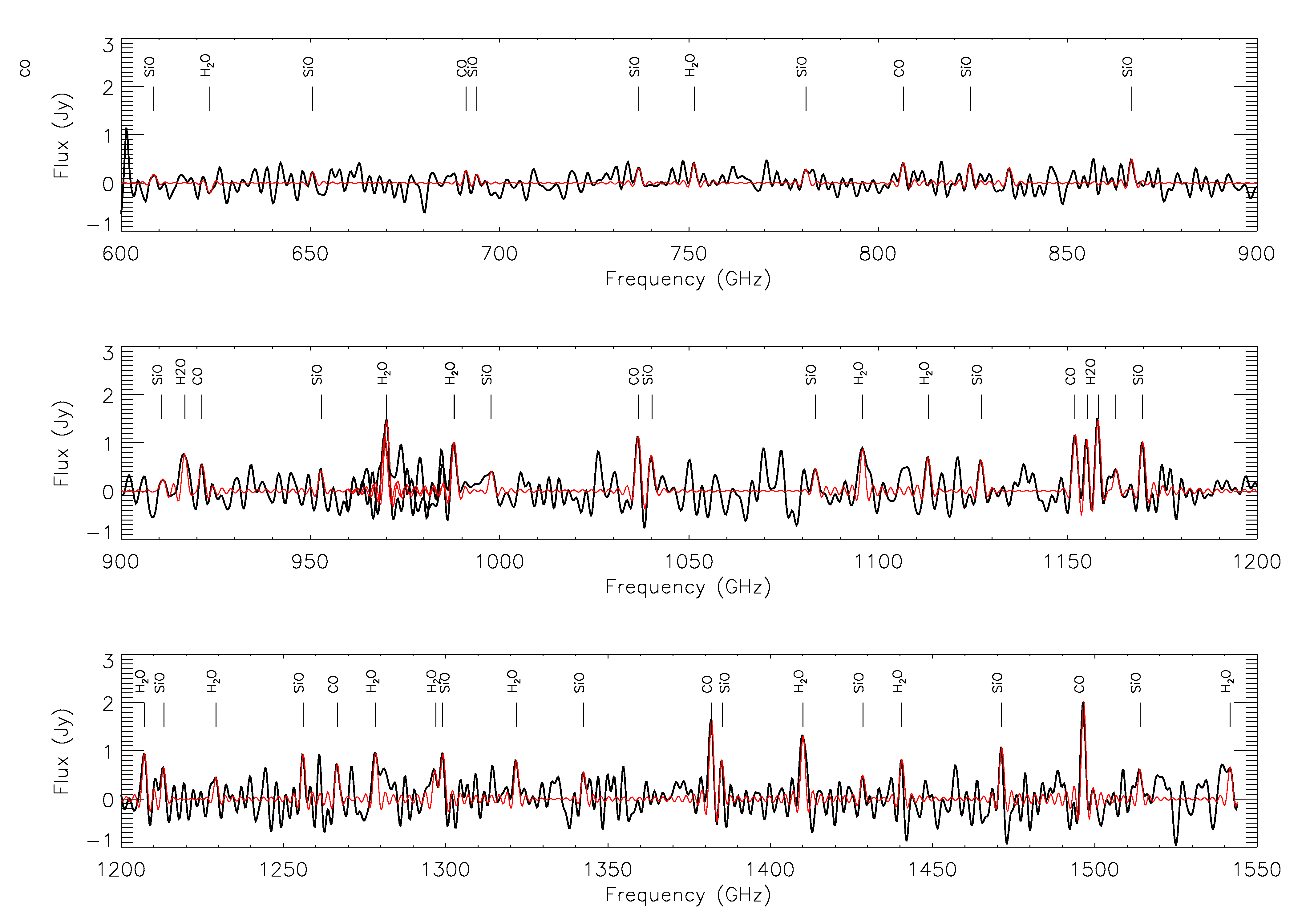}
   \caption{Rectified  SPIRE spectrum (unapodised) from 600 GHz to 1550 GHz (black) overlaid with the  fitted spectrum of the three major species CO, SiO and H$_2$O (red). We note that this spectrum does not include the scaling for the flux losses due to off-centring (see text).}
  \label{fig:sspec}
\end{figure*}

We reduced the SPIRE spectra using the standard point-source pipeline described by \citet{Fulton:2010gt} using the HIPE calibration scheme 12\_1 described by \citet{Swinyard:2014vy}.  The size of the SPIRE beam (17\arcsec--40\arcsec\ depending on frequency--see  \citealt{Makiwa:2013ho})  is such that some background emission will be captured as well as the source itself (see App.\,\ref{sec:app3}).  A careful inspection of the spectra in the surrounding bolometers showed no evidence of strong emission lines and we only removed a polynomial fit to the continuum (to subtract {\sl all} continuum sources present), resulting in the rectified spectrum illustrated in Fig.\,\ref{fig:sspec}.  

 This continuum subtraction adds an extra uncertainty to the subsequent spectral fitting, since our source is faint and the continuum noisy. 
We estimate an uncertainty in the continuum placement of $\pm0.1$\,Jy. The effect of this is partly accounted for in the fitting errors:  a sinc profile placed too low below 0 will be fit worse than a sinc correctly set to the zero level. We have, however, also added this uncertainty, converted to an integrated flux ($1.2\times10^{-18}$\,W\,m$^{-2}$), to the errors taken into account in our modelling of the integrated line fluxes (Sec.\,\ref{sec:cosio}).

As with PACS, V838\,Mon is offset from the centre of the FoV. Comparing the location of the star on the PACS and SPIRE images and the location of the central SPIRE bolometers, we determined an offset of close to 6\arcsec. This has consequence for the point-source calibration, which does not account for the flux loss due to off-centring. We used a SPIRE script provided in HIPE (Track 14) to calculate the underestimate of the fluxes due to this off-centring: the loss factor is about 1.04 below 1000\,GHz and from 1.25 to 1.40 above 1000GHz. We note that  this task and the correction factors were available only after we had reduced and measured our spectra, but all fluxes and calculations presented here  do include this offset correction, unless otherwise stated.

The SPIRE FTS calibration is  based on the spectrum of Uranus and detailed analysis of the flux uncertainties \citep{Swinyard:2014vy} which shows that the absolute flux on a source of this brightness is calibrated to within 6\%.

\subsection{HIFI spectroscopy}\label{sec:hifispec}

We acquired five double sideband pointing observations with HIFI, further referred to as settings A-E. Details are given in Table~\ref{tab:obs}.  Settings A (in band 1b), B (in band 2a), and C (in band 5a) were observed in the standard dual beam switch (DBS) mode, while the FastDBS mode was used for the high-frequency settings D (in band 7a) and E (in band 7b) in order to optimise baseline stability. Data were obtained at nominal resolution for both the high-resolution spectrometer (HRS) and the wideband spectrometer (WBS), in both the horizontal and vertical polarisations. The bandwidth of the HRS is narrow, at 230\,MHz, that is 115\,km\,s$^{-1}$ and 40\,km\,s$^{-1}$ in settings A and E, respectively. 
The antenna's half power beamwidth is 34\arcsec.8, 31\arcsec.2, 18\arcsec.4, 12\arcsec.0, and 11\arcsec.9, for settings A through E \citep[][see their eq.~3]{2012A&A...537A..17R}. The HIFI data obtained with the WBS were reduced up to level two using the pipeline available within HIPE.  Baseline subtraction and standing-wave removal in the WBS spectra were severely hampered by the noise levels present in the data. Additionally, the lack of coinciding features between the data obtained in the vertical and horizontal polarisations led us to conclude that the HIFI spectra yielded no reliable emission line measurements. This is likely a consequence of the lines being too faint and/or too wide. The ortho-H$_{2}$O $7_{3,4}-7_{2,5}$ transition at 166.8\,$\mu$m (1797.159\,GHz), detected in the PACS observations, is covered by setting E, but no emission feature is seen in either polarisation. Using the point source sensitivities for HIFI, listed by  \citet{2012A&A...537A..17R}, we find that the flux measured by PACS can indeed not be retrieved at the sensitivities reached in our HIFI observations. Moreover, \citet{2009A&A...503..899T} measured an outflow with a terminal velocity of 215\,km\,s$^{-1}$ from numerous lines from V838\,Mon, giving a total linewidth of $\sim2580$\,MHz at 1797.159\,GHz. Considering the WBS bandwidth of 2.56\,GHz in setting E, it is clear that if this lines were indeed this broad, we would not only struggle to see it but also to reduce the data in a reliable way.

\section{V838\,Mon: point source photometry}\label{sec:photmeasp}

As is clear from Fig.\,\ref{fig:firstmaps}, the emission on the maps consists of a bright point source (V838\,Mon itself) and fairly bright extended emission surrounding the star; for SPIRE the two emission sources are particularly intimate. To measure the flux density from the star, aperture photometry using various tasks in HIPE was possible for the PACS data, but for  SPIRE  the relative strength and unevenness of the extended emission made such measurements highly uncertain. We therefore took an alternative approach for SPIRE, which we also tested for PACS: to scale the PSF (point spread function; beam) and subtract it from the  maps until only residual emission remained at the position of the star. The photometry of the scaled beam is then the photometry of V838\,Mon. As a refinement of this method, we also performed PSF-subtraction on deconvolved SPIRE maps. Although the results were not as useful as hoped, we believe a summary of this work will be useful for others (App.\,\ref{sec:decon}).

Jumping ahead to the final results for the point source (Sec\,\ref{subsec:sed}): within the 2$\sigma$ errors there is no change in the point-source photometry between the three epochs, which cover 16 months. The properties of the dust giving rise to this emission have not changed noticeably during this period.

\subsection{PACS}\label{sec:photmeaspp}

For the aperture photometry we adopted aperture sizes narrow enough to avoid including too much extended emission, and adopted the appropriate aperture and colour corrections: see App.~\ref{sec:app1} for the details.  The errors in the photometry arising from the scatter in the sky background on the maps are 0.02\,Jy (70\,$\mu$m), 0.04\,Jy (100\,$\mu$m), and  0.12\,Jy (160\,$\mu$m), and these values were folded into the errors arising from the photometric measurements. Another source of uncertainty comes from the contaminating flux from the extended emission at the position of the star, which while not strong -- the ratio of the peak flux from the star to that in the immediately surrounding extended emission is  of the order 40-30-5 for the blue-green-red maps -- it is non-negligable. To account for this, we also measured the flux values from the extended emission  close to the position of the star and consider this to be the sky flux to be subtracted from the aperture photometry. The photometry given in Table\,\ref{tab:pphot} is then mean$\pm$range of these two sets of measurements.

We also performed PSF-subtraction on the PACS maps of V838\,Mon, using maps of the PSF taken from the PACS calibration web-page: see below and App.~\ref{sec:app5} for more detail. The main reason for doing this was to obtain maps of the extended emission only, but the method can also be used to obtain photometry for V838\,Mon.  Details of the photometric corrections (aperture and colour corrections) made on the scaled PSF are given in  App.~\ref{sec:app1}, and the method itself is explained in App.~\ref{sec:app5}.  The photometry resulting from this are  also given  in Table\,\ref{tab:pphot}. We find V838\,Mon to have a FWHM consistent with that of the PSF source we used, but with a 2d profile that is somewhat different: residuals in the subtraction of the point source from V838\,Mon indicate a difference in the core and the wings of the profile (see App.~\ref{sec:app5}). This is the main contributor to the photometry errors arising from this process. 

The PSF-subtracted maps were used in the study of the extended emission. After subtracting out the point source, the region left behind (which contains the above-mentioned residuals) was replaced by a circle of constant flux (value taken from the local background). 

\subsection{SPIRE}\label{sec:photmeasps}

To measure the photometry of V838\,Mon from the SPIRE maps we used a PSF-subtraction method. Aperture photometry was not feasible because of the relative brightness of the extended source: indeed at 500\,$\mu$m the extended source almost completely overwhelms the point source (see e.g. Fig.\,\ref{fig:resid}). We used the maps produced in HIPE and calibrated for point sources, that is with units of Jy/beam, to obtain the point-source photometry, but performed the same procedure on the maps calibrated for extended sources to create point-source-free maps of the extended emission.

The PSF-subtraction method is essentially a scale-and-subtract, and the details of this work, which was carried out in HIPE, can be found in App.~\ref{sec:app5}. The photometry is then measured from the beam maps which are scaled sufficiently to remove the point source emission from the V838\,Mon maps when image subtraction is done. We used a SPIRE ``useful'' photometry script in HIPE for the photometry measurements, using the source fitting photometry task ({\it sourceExtractorSussextractor}), which works directly on the point-source calibrated maps. See App.~\ref{sec:app1} for details of the photometric corrections. 
The errors in the photometry with this method are dominated by the  range of  values that result in acceptable PSF-subtraction residuals, and for the 500\,$\mu$m maps are especially large as the beam is large compared to the size of the extended emission; here we add that the result from our PSF subtraction from the deconvolved maps (App.\,\ref{sec:decon}) is that the fluxes at this wavelength may be slightly overestimated.  The values and errors given in Table\,\ref{tab:pphot} are taken from the acceptable range of PSF scaling factor values. 

After subtracting out the point source, the region left behind was replaced by a circle of constant flux (value taken from the local background). These cleaner PSF maps were then used to study the extended emission.

\begin{table*}[ht!]
\caption{PACS and SPIRE point and extended source spectro-/photometric measurements of V838\,Mon (except for those not applicable ``na''). All values have been colour, aperture, and beam corrected (see the text). One-sigma measurement errors including the uncertainty due to the background subtraction are given. The calibration uncertainties (as well as other sources of error) are discussed in the text, and are 5\% for PACS and 6\% for SPIRE photometry. 
Spectroscopy values are given for comparison to the more accurate photometry.}
\label{tab:pphot}
\centering
\begin{tabular}{rcccp{3mm}ccc}\hline
\multicolumn{4}{c}{Point source}                        &     & \multicolumn{3}{c}{Extended source (point-source-free)} \\ \cline{1-4}\cline{6-8}
Band,  & Aperture       & PSF           &  Spect.\tablefootmark{a}       &     & Flux density    & Area  & Size\tablefootmark{b} \\
epoch  & photometry     & subtraction   &  cont.        &     &                  &       & \\ 
       & Jy             & Jy            &  Jy           &     & Jy              & $^\square$\arcsec  &  $\arcsec$ \\\hline
70, 1  & $7.33\pm0.12$  & 6.8--7.4      &  $6.0$  &     & $ 3.26 \pm0.12 $& 8860  & $90\times110$   \\
70, 2  & $7.60\pm0.12$  & 7.0--7.4      & -             &     & $ 2.93 \pm0.12 $& 8530  & \\                
70, 3  & $7.54\pm0.11$  & 6.8--7.2      & -             &     & $ 2.62 \pm0.11 $& 7270  & \\                
100, 1 & $3.17\pm0.07$  & 3.2--3.6      &  $3.0$  &     & $10.51 \pm0.07 $& 13390 & $130\times100$ \\
100, 2 & $3.38\pm0.04$  & 3.5--3.7      & -             &     & $11.00 \pm0.04 $& 11400 & \\               
100, 3 & $3.35\pm0.06$  & 3.5--3.7      & -             &     & $10.53 \pm0.06 $& 12760 & \\               
160, 1 & $1.07\pm0.15$  & 1.1--1.5      & $2.0$   &     & $14.25 \pm0.15 $& 14250 & $140\times110$  \\
160, 2 & $1.17\pm0.15$  & 1.2--1.5      & -             &     & $14.17 \pm0.15 $& 14340 & \\                
160, 3 & $1.16\pm0.15$  & 1.1--1.4      & -             &     & $14.97 \pm0.15 $& 16820 & \\                
250, 1 & na             & $0.56\pm0.10$ & $\sim0.7$     &     & $ 6.78 \pm0.10 $& 20100 & $160\times120$\\  
250, 2 & na             & $0.59\pm0.06$ & -             &     & $ 6.15 \pm0.07 $& 17400 & \\                
250, 3 & na             & $0.59\pm0.08$ & -             &     & $ 5.82 \pm0.08 $& 17600 & \\                
350, 1 & na             & $0.27\pm0.06$ & $\sim0.3$     &     & $ 3.08 \pm0.07 $& 20100 & $140\times140$\\ 
350, 2 & na             & $0.28\pm0.05$ & -             &     & $ 2.21 \pm0.06 $& 14600 & \\                
350, 3 & na             & $0.31\pm0.06$ & -             &     & $ 2.70 \pm0.07 $& 18500 & \\                
500, 1 & na             & $0.20\pm0.05$ & $\sim0.1$     &     & $ 0.82 \pm0.05 $& 17700 & $160\times130$\\ 
500, 2 & na             & $0.19\pm0.06$ & -             &     & $ 0.72 \pm0.06 $& 15400 & \\                
500, 3 & na             & $0.19\pm0.07$ & -             &     & $ 1.00 \pm0.07 $& 21000 & \\          
\hline
\end{tabular}
\tablefoot{
\tablefoottext{a}{Measured from the spectral segment closest to the photometry wavelength.}
\tablefoottext{b}{Mean diameter at position angles $\pm45$\degr.}
}
\end{table*}

\section{V838\,Mon: extended source photometry}\label{sec:photmease}

For PACS, the sum of the flux in all the pixels within a certain contour on the maps,  including the point source, was measured. Subtracting the flux of the point source then gave the extended source flux density values. The contours were chosen for each map to encompass the flux close to the level of the sky, with a secondary consideration given to having the area similar for each epoch: the contours values taken lie between 2$\sigma$ and 4$\sigma$ above the  mean background  value on the residual maps, where $\sigma$ is the standard deviation of the sky flux as measured from about 10 apertures. 

\begin{figure*}
  \centering
  \includegraphics[scale=0.48]{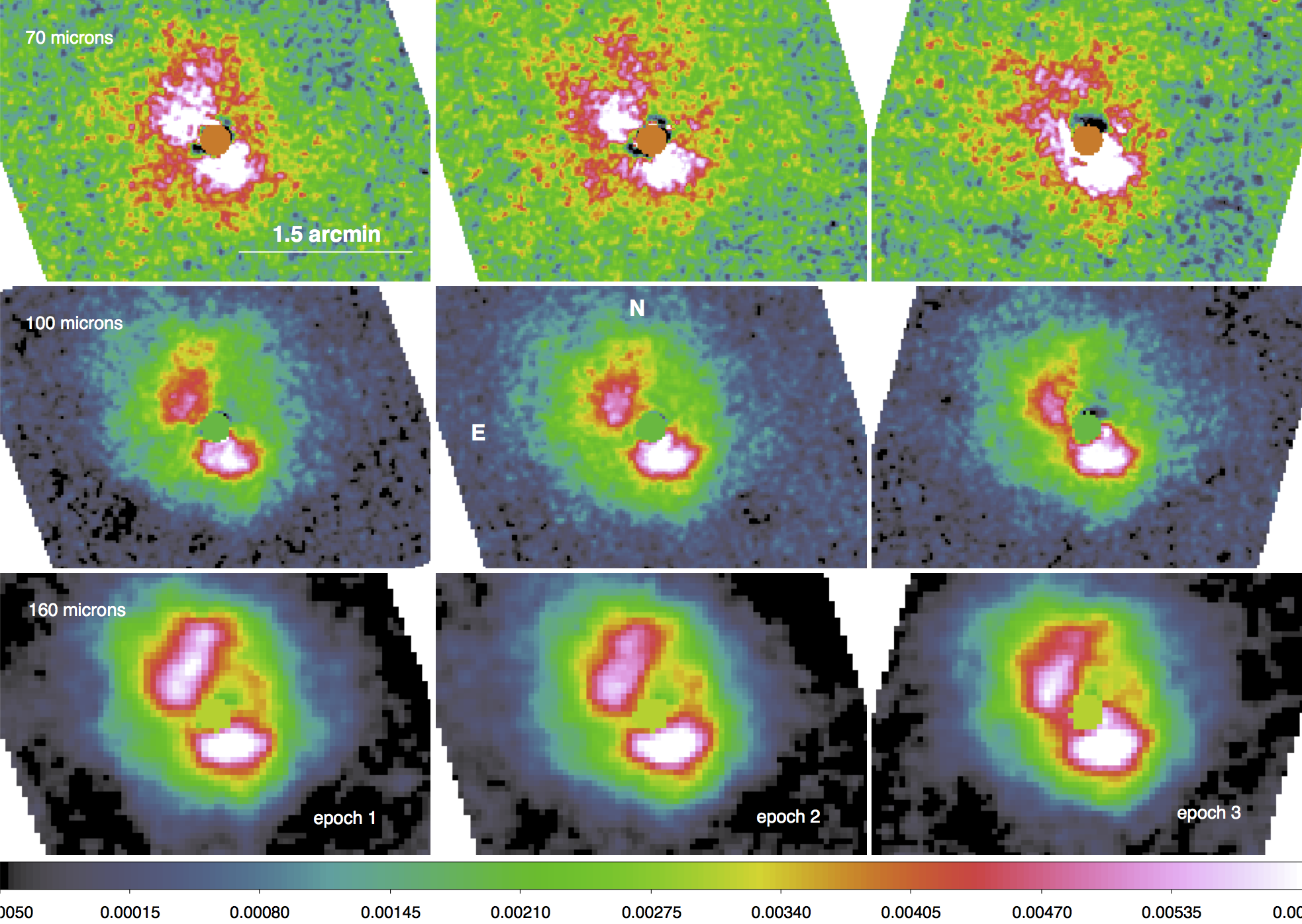}
  \includegraphics[scale=0.48]{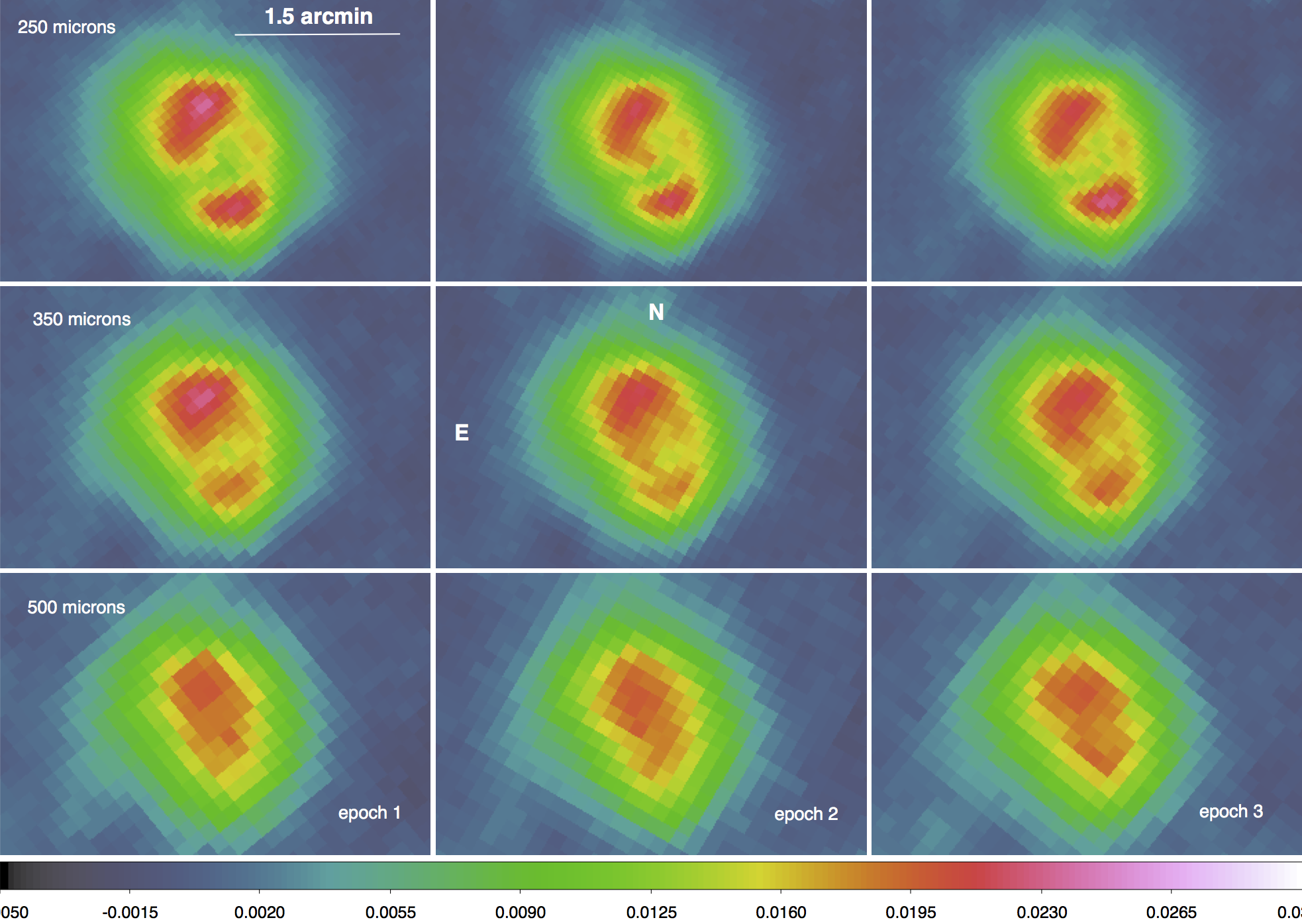}
   \caption{SPIRE and PACS maps with the point source removed and residual replaced by a blank patch (on the 70$\mu$m  some residuals remain). Scaling ranges are chosen based on the peak flux density. PACS: -0.0005--0.001 Jy/pixel (70\,$\mu$m), -0.0004--0.006 Jy/pixel (100\,$\mu$m), -0.0001--0.02 Jy/pixel (160\,$\mu$m), with pixel sizes of 1\arcsec, 1.5\arcsec, and 3\arcsec, respectively. SPIRE: -0.005--0.03 Jy/pixel (250\,$\mu$m), -0.005--0.025 Jy/pixel (350\,$\mu$m),  -0.005--0.02 Jy/pixel (500\,$\mu$m), with pixels sizes of 4.5\arcsec, 6.25\arcsec, and 9\arcsec, respectively. N--E is indicated in the central maps.}
  \label{fig:pext}
\end{figure*}

The SPIRE maps used were those calibrated for extended emission (units MJy/sr). From these measurements we subtracted the point-source photometric measurements  to obtain the photometry of the extended emission only. We compared these values to the values obtained from measuring the photometry of the extended source from the PSF-subtracted maps (Sec.\,\ref{sec:photmeasps}: the extended emission with the point source subtracted): these agree to better than $\pm5$\%. We then applied the extended-source colour corrections and beam-area corrections to the result: see App.~\ref{sec:app1} for details of the corrections and Table\,\ref{tab:pphot} for the measurements. The error for the fluxes of the extended source is dominated by the error in the point-source photometry. Calibration uncertainties are not included in the error values. The PSF-subtracted maps (SPIRE and PACS) are shown in Fig.\,\ref{fig:pext}.

The photometric results given in Table\,\ref{tab:pphot} show that there is no difference, within generally less than 2$\sigma$, in the point source fluxes from epoch to epoch. For the extended source, from looking at the maps of Fig.\,\ref{fig:pext} and at the better detail in the point-source subtracted  deconvolved maps (App.\,\ref{sec:decon}), it does seem that the flux in the northern part of the source decreases with time compared to the flux in the southern part of the source, at all wavelengths (excepting perhaps 500\,$\mu$m). However, these and any other differences between epochs are mostly within 3$\sigma$, and the derived dust properties do not change significantly with epoch (c.f. Sec.\,\ref{subsec:sed}). 

\section{V838\,Mon: point source spectroscopy}\label{sec:specmeas}

\subsection{PACS}\label{sec:pacsspecmeas}

The V838\,Mon PACS spectra were measured in HIPE, fitting a low-order polynomial and a Gaussian to all features with the width and peak intensity to appear to be emission lines. We then compared the line list to that from the red supergiant VY\,CMa: this star has a well-studied PACS--SPIRE spectrum, and it has a cool and extended atmosphere \citep{2010A&A...518L.145R, Matsuura:2013cz}, as V838\,Mon is suggested to have. The comparison was used as a guide for a second pass over the spectra to look for additional features and eliminate those most likely to be noise. Most of the lines we identified this way are of water. The PACS spectra have units of Jy and $\mu$m; the line fluxes we report are the integrated intensity of the fitted Gaussian in units of W\,m$^{-2}$ (computed as $peak\times\sigma\times(2\pi)^{0.5}\times3\times10^{-12}/\lambda^2$, where $\sigma$ is the width of the Gaussian and the last phrase is to convert the units). Measurement errors were propagated according to the standard rules. Table\,3 gives the positions, fluxes and uncertainties of all the lines identified in the PACS (and SPIRE) spectra.   
The lines for which we could not assign a clear identification are given in Table\,\ref{tab:unid}.
 
The PACS lines are redshifted by $\sim0.02$--0.03\,$\mu$m (blue to red wavelengths) compared to our identifications, corresponding to 60--80\,km\,s$^{-1}$. 
The PACS pipeline produces wavelengths corrected to LSR: converting to heliocentric produced redshifts of 40--60\,km\,s$^{-1}$
which is close to previous redshift reported values 58\,km\,s$^{-1}$ \citep{2009A&A...503..899T} or  71\,km\,s$^{-1}$ \citep{2011A&A...532A.138T}. Some of our measured shift, however,  will be caused by V838 Mon being located slightly off from the centre of the central spaxel (PACS Observer's Manual\footnote{see http://herschel.esac.esa.int/twiki/pub/Public/PacsCalibrationWeb}). V838\,Mon lies between the central spaxel (12) and its neighbour (17) by less than 1/2 spaxel, and this will cause the spectrum obtained from the central spaxel to have a redshift of  $\sim0.1$--0.2\,$\mu$m (blue to red wavelengths) -- although this effect is diluted somewhat since our final spectrum is a combination of spaxel 12 (slight redshift) and 17 (slight blueshift).

\begin{table*}[ht!]
\caption{PACS and SPIRE line fluxes and errors: 1-$\sigma$ signal-to-noise ratio for the SPIRE lines and the measurement error for the PACS lines. The calibration uncertainties are: PACS: 8(blue)--12(red)\% SPIRE: 6\%\ with an additional $1.2\times10^{-18}$\,W\,m$^{-2}$ arising from the continuum rectification. The PACS lines start at 1660\,GHz.  All lines in our line list are given here, including those for which the error is larger than the value. 
}
\label{tab:lines}
\centering
\begin{tabular}{lccccc}\hline
Species	&	Observed $\nu$ (GHz)	&	Rest $\nu$ (GHz)	&	$\delta\nu$ (GHz)	&	Flux\tablefootmark{a}
 	&	$\delta$F\tablefootmark{a}  	\\\hline

 H$_2$O $4_{2,3}$--$3_{3,0}$	\tablefootmark{b}&	448.63	&	448.001	&	0.04	&	1.16	&	0.15			\\
CO 4--3\tablefootmark{c}						&	461.19	&	461.041	&	0.21	&	0.46	&	0.14			\\
SiO 11--10					&	478.10	&	477.505	&	0.17	&	0.62	&	0.14			\\
H$_2$O $1_{1,0}$--$1_{0,1}$	&	557.36	&	556.936	&	0.14	&	0.34	&	0.14			\\
CO 5--4						&	575.52	&	576.268	&	0.32	&	0.30	&	0.14			\\
H$_2$O $5_{3,2}$--$4_{4,1}$	&	621.41	&	620.701	&	0.41	&	0.11	&	0.14			\\
CO 6--5						&	691.09	&	691.473	&	0.30	&	0.32	&	0.14			\\
H$_2$O $2_{1,1}$--$2_{0,2}$	&	751.33	&	752.033	&	0.12	&	0.45	&	0.14			\\
CO 7--6						&	806.65	&	806.652	&	0.21	&	0.46	&	0.14			\\
SiO 19--18					&	824.28	&	824.236	&	0.20	&	0.50	&	0.14			\\
SiO 20--19					&	866.88	&	867.523	&	0.16	&	0.64	&	0.14			\\
H$_2$O $4_{2,2}$--$3_{3,1}$	&	916.52	&	916.172	&	0.06	&	0.97	&	0.14			\\
CO 8--7						&	921.18	&	921.800	&	0.16	&	0.60	&	0.14			\\
SiO 22--21					&	952.78	&	954.054	&	0.21	&	0.47	&	0.14			\\
H$_2$O $5_{2,4}$--$4_{3,1}$\tablefootmark{d}	&	969.64	&	970.315	&	0.15	&	1.40	&	0.14			\\
H$_2$O $5_{2,4}$--$4_{3,1}$\tablefootmark{e}	&	970.35	&	970.315	&	0.07	&	1.37	&	0.16			\\
H$_2$O $2_{0,2}$--$1_{1,1}$\tablefootmark{d}	&	987.96	&	987.927	&	0.14	&	1.30	&	0.15			\\
H$_2$O $2_{0,2}$--$1_{1,1}$\tablefootmark{e}	&	988.12	&	987.927	&	0.09	&	1.07	&	0.16		\\
SiO 23--22					&	997.63	&	997.297	&	0.40	&	0.57	&	0.15			\\
CO	9--8					&	1036.62	&	1036.912	&	0.29	&	1.02	&	0.16			\\
SiO 24--23					&	1040.36	&	1040.523	&	0.27	&	0.98	&	0.15			\\
H$_2$O $3_{1,2}$--$3_{0,3}$	&	1095.79	&	1097.365	&	0.11	&	1.04	&	0.16			\\
H$_2$O $1_{1,1}$--$0_{0,0}$	&	1113.01	&	1113.343	&	0.12	&	0.87	&	0.16			\\
SiO 26--25					&	1126.84	&	1126.922	&	0.25	&	0.99	&	0.15			\\
H$_2$O $3_{1,2}$--$2_{2,1}$\tablefootmark{g}	&	1151.92	&	1153.127	&	0.06	&	1.65	&	0.16			\\
H$_2$O $6_{3,4}$--$5_{4,1}$	&	1158.11	&	1158.324	&	0.22	&	1.61	&	0.16			\\
H$_2$O $3_{2,1}$--$3_{1,2}$	&	1163.14	&	1162.912	&	0.34	&	0.84	&	0.16			\\
SiO 27--26					&	1169.84	&	1170.094	&	0.31	&	0.77	&	0.15			\\
H$_2$O $4_{2,2}$--$4_{1,3}$	&	1207.39	&	1207.639	&	0.21	&	1.25	&	0.16			\\
SiO 28--27					&	1213.16	&	1213.247	&	0.27	&	0.85	&	0.15			\\
H$_2$O $2_{2,0}$--$2_{1,1}$	&	1228.68	&	1228.789	&	0.27	&	0.82	&	0.16			\\
SiO 29--28					&	1256.22	&	1256.380	&	0.18	&	1.24	&	0.15			\\
CO	11--10					&	1266.78	&	1267.014	&	0.18	&	1.37	&	0.16			\\
H$_2$O $7_{4,3}$--$6_{5,2}$	&	1278.16	&	1278.266	&	0.17	&	1.09	&	0.16			\\
H$_2$O $8_{2,7}$--$7_{3,4}$	&	1297.16	&	1296.411	&	0.21	&	0.86	&	0.16			\\
SiO 30--29					&	1298.88	&	1299.491	&	0.23	&	0.98	&	0.15			\\
H$_2$O $6_{2,5}$--$5_{3,2}$	&	1321.77	&	1322.065	&	0.14	&	1.30	&	0.16			\\
SiO 31--30					&	1342.40	&	1342.582	&	0.29	&	0.76	&	0.15			\\
CO 12--11					&	1381.96	&	1381.995	&	0.10	&	2.29	&	0.16			\\
SiO 32--31					&	1385.15	&	1385.650	&	0.21	&	1.02	&	0.15			\\
H$_2$O $5_{2,3}$--$5_{1,4}$	&	1410.00	&	1410.618	&	0.11	&	1.66	&	0.16			\\
H$_2$O $7_{2,6}$--$6_{3,3}$	&	1440.62	&	1440.782	&	0.15	&	1.18	&	0.16			\\
SiO 34--33					&	1471.38	&	1471.717	&	0.13	&	1.63	&	0.15			\\
CO 13--12					&	1496.56	&	1496.923	&	0.08	&	2.90	&	0.16			\\
SiO 35--34					&	1513.57	&	1514.714	&	0.21	&	1.06	&	0.15			\\
H$_2$O $6_{3,3}$--$5_{4,2}$	&	1541.72	&	1541.967	&	0.21	&	0.96	&	0.16			\\
 \hline
\end{tabular}
\tablefoot{
\tablefoottext{a}{$10^{-17}$ W\,m$^{-2}$}
\tablefoottext{b}{$J'_{{K'_a},{K'_c}}$--$J_{{K_a},{K_c}}$}
\tablefoottext{c}{$J'$--$J$}
\tablefoottext{d}{Seen in SLW Detector}
\tablefoottext{e}{Seen in SSW Detector}
\tablefoottext{g}{Blended with CO 10--9}
}
\end{table*}

 \addtocounter{table}{-1}
\begin{table*}
\caption{-- continued.}
\centering
\begin{tabular}{lccccc}\hline
Species	&	Observed $\nu$ (GHz)	&	Rest $\nu$ (GHz)	&	$\delta\nu$ (GHz)	&	Flux\tablefootmark{a}
 	&	$\delta$F\tablefootmark{a}  	\\\hline
H$_2$O $2_{2,1}$--$2_{1,2}$	&	1660.52	&	1661.008	&	0.09	&	2.58	&	0.89			\\
H$_2$O $2_{1,2}$--$1_{0,1}$	&	1669.37	&	1669.905	&	0.09	&	3.64	&	0.78			\\
SiO 39--38					&	1686.52	&	1686.444	&	0.09	&	1.46	&	0.48			\\
H$_2$O $6_{3,3}$--$6_{2,4}$	&	1761.79	&	1762.043	&	0.10	&	5.79	&	0.73			\\
H$_2$O $7_{3,5}$--$6_{4,2}$	&	1765.87	&	1766.199	&	0.10	&	2.53	&	0.70			\\
SiO 41--40					&	1771.60	&	1772.144	&	0.10	&	0.63	&	0.70			\\
H$_2$O $7_{3,4}$--$7_{2,5}$	&	1797.05	&	1797.159	&	0.11	&	2.38	&	0.37			\\
SiO 42--41					&	1814.33	&	1814.950	&	0.11	&	1.08	&	0.48			\\
CO 16--15					&	1840.98	&	1841.346	&	0.11	&	2.82	&	0.45			\\
H$_2$O $3_{3,1}$--$4_{0,4}$	&	1893.45	&	1893.687	&	0.12	&	4.67	&	0.17			\\
H$_2$O $6_{4,3}$--$6_{3,4}$	&	2772.56	&	2773.977	&	0.26	&	6.79	&	1.31			\\
H$_2$O $7_{4,4}$--$7_{3,5}$	&	3329.96	&	3329.185	&	0.37	&	13.10	&	1.20			\\
H$_2$O $7_{1,6}$--$7_{0,7}$	&	3535.54	&	3536.667	&	0.42	&	9.31	&	0.89			\\
H$_2$O $6_{0,6}$--$5_{1,5}$	&	3599.04	&	3599.642	&	0.65	&	5.07	&	0.88			\\
H$_2$O $9_{2,7}$--$9_{1,8}$	&	3681.83	&	3682.708	&	0.68	&	8.68	&	0.76			\\
H$_2$O $7_{2,6}$--$7_{1,7}$	&	3690.75	&	3691.316	&	0.68	&	5.67	&	0.70			\\
H$_2$O $5_{2,4}$--$4_{1,3}$	&	4217.43	&	4218.431	&	0.89	&	8.42	&	2.98			\\
H$_2$O $8_{2,7}$--$8_{1,8}$	&	4240.27	&	4240.192	&	0.90	&	8.16	&	4.18			\\
H$_2$O $5_{3,2}$--$5_{0,5}$	&	5499.40	&	5500.117	&	1.51	&	24.50	&	4.40			\\
\hline
\end{tabular}
\tablefoot{
\tablefoottext{a}{$10^{-17}$ W\,m$^{-2}$}
}
\end{table*}

In Table\,\ref{tab:pphot} we have indicated the approximate values of the spectral continuum flux at the wavelengths of the photometric bands.  The spectroscopy and photometry match within their respective uncertainties. 

\subsection{SPIRE}\label{sec:spirespecmeas}

Fourier transform spectrometers produce spectra in which lines have sinc profile functions \citep{2014SPIE.9143E..2DN} that are linearly sampled in frequency.  Therefore, all our FTS spectral measurements have been made in frequency space and we have retained the natural sinc profile for all spectral measurements.  The integrated line strengths were obtained using a built-in HIPE line fitting script that fits a sinc profile to a given line position using least square fitting of the instrumental line profile \citep{2014SPIE.9143E..2DN}. The uncertainty in the line is taken from the standard deviation of the residuals around the line after subtraction of the fitted profile, combined with the line fitting errors.  Table\,3 gives the positions, fluxes and uncertainties of all the lines identified in the SPIRE (and PACS) spectra.  

Because our analysis focuses on CO, SiO, and H$_2$O, we fit only these lines in the SPIRE spectra.
All of these species have clear lines present, in both the PACS and SPIRE spectral ranges.  Our line-list fitting to the SPIRE spectra is shown in Fig.\,\ref{fig:sspec}. We note that there are  features clearly visible in the spectrum which we have not fit. Some of these could be emission lines that were not on our line list, while others will be noise.  An advantage of fitting a sinc profile is that it is not just the main peak that needs to fit the data, but the minor peaks in the wings also. The fitting errors are based on the fit to the entire profile, and hence give a more reliable estimate than if fitting the SPIRE data apodised to a Gaussian profile.  However, a disadvantage of spectra with a sinc profile is that the multiple-peak nature of this profile leads to some noise spikes being as bright as real lines. It is for this reason that we fit the SPIRE spectra with a line list: we know the exact position of the expected lines and fit only those. All lines from the line list are included in Table 3, and where the fitting errors exceed the integrated flux, the measured flux is clearly to be considered unreliable.

\section{Analysis}\label{sec:analysis}
\subsection{Far infrared spectral energy distribution: (dust) continuum emission}\label{subsec:sed}

The total integrated \emph{Herschel} flux density for the central (point source) emission and the large-scale extended infrared emission, for each epoch, are plotted in Fig.\,5.  
In addition, the full SPIRE spectrum of the central point source (smoothed) is shown in light grey.
The near- to far-infrared spectral energy distribution of the central (point source) is shown in Fig.\,6 ({\sl Herschel}-only and with literature data added). The fitting results are given in Tables 4 and 5.

\paragraph{Cool extended source}
A modified black body was fitted to the total integrated extended far-infrared \emph{Herschel} emission (including data from all epochs) given in Table\,\ref{tab:pphot}. The emitted flux density is modelled with the functional form: $S_\nu \propto\ \kappa_\nu\ B_\nu$, with $\kappa_\nu = 0.1\ (\nu/1000\ \mathrm{GHz})^\beta = 0.1\ (300 \mu\mathrm{m}/\lambda)^\beta$~cm$^2$/g (c.f. Hildebrand 1983; Beckwith et al.\ 1990), assuming a gas-to-dust ratio of 100. The power-law index $\beta$ is an indicator of dust grain composition. $\beta \sim 2$ for silicate dust and $\beta \sim 1$ for amorphous carbon dust \citep{Mennella:1998p29412, Boudet:2005hn}. For interstellar dust $\beta$ ranges from 1.7--2.6 \citep{Paradis:2010p29033, Gordon:2014by}. For this cool material the \emph{Herschel} wavelength coverage provides a good constraint to the dust temperature, T$_\mathrm{dust}$, dust mass, M$_\mathrm{dust}$, and the emissivity index, $\beta$. The best-fit result is shown in Fig.\,5 and values for the fitted parameters for the extended emission are labelled in the figure. 
\begin{figure}[t!]
\begin{center}
\includegraphics[height=6cm]{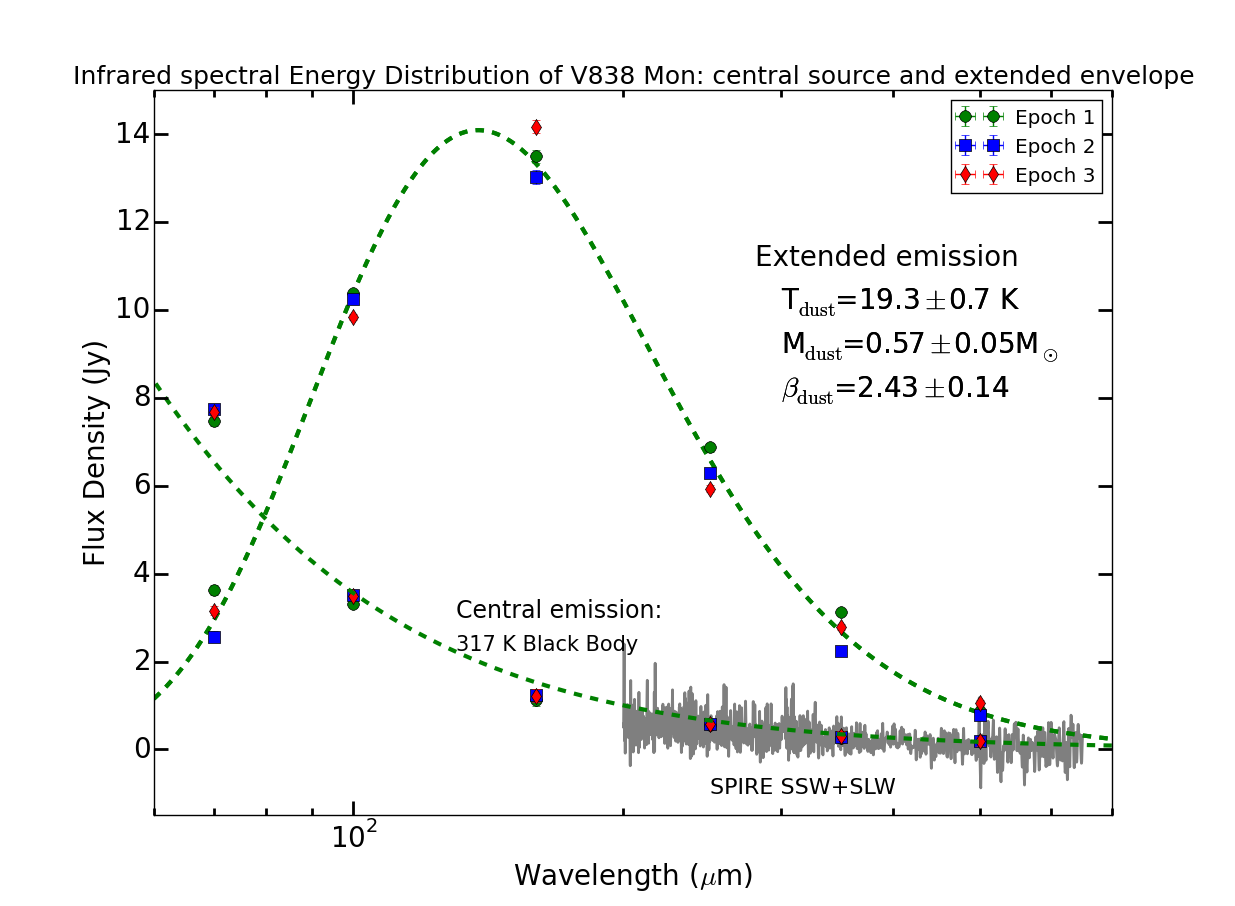}
\label{fig:sedherschel}
\caption{Spectral energy distribution (SED) of the unresolved central point source and the extended far infrared emission observed with \emph{Herschel} PACS and SPIRE. 
The blackbody curve shown for the central (point source) emission is derived from the near- to far-infrared SED shown in the right panel.
The modified black body shown for the extended emission is the best-fit to the \emph{Herschel} measurements (averaged over all epochs).}
\end{center}
\end{figure}

The dust temperature for the extended source as derived from \emph{Herschel} data is T$_\mathrm{dust} = 19.3 \pm 0.7$~K, which is close to the value of 25$^{+4}_{-5}$ derived by \citet{2006ApJ...644L..57B} from 2005 Spitzer--MIPS photometry. The fitted dust emissivity index, $\beta$, is 2.43$\pm$0.14, which fits into the range of that for ISM dust.  The required cold dust mass is 0.57 $\pm$ 0.05 M$_\odot$ (at 6.2\,kpc). 

\citet{2006ApJ...644L..57B} derived a value of  0.9 M$_\odot$ (converted to a distance of 6.2\,kpc). Their higher mass value is because of their higher measured flux densities for the extended emission taken from {\it Spitzer} data at 70 and 160~$\mu$m from 2004/5. The difference ($\sim11$\,Jy at 70\,$\mu$m compared to our $\sim3$\,Jy) is quite striking. However, our derived mass is sufficiently large that the conclusion of Banerjee et al.\ -- that it is ISM material -- still holds. We discuss this further in Sec.\,\ref{sec:spitzercomp}.

Although there is some scatter in the flux densities derived for each epoch, for both the extended source and the point source (see below), there is no evidence for significant variation in the total integrated spectral energy distribution, and the fits to the individual epochs always yield consistent result, as shown in Table 4.

\begin{table}
\begin{center}
\label{tb:sedfitextendedherschel}
\caption{{\bf Upper}: ODR fitting (scipy.odr) results of \emph{Herschel} SED of the
extended infrared emission (Sec.\,\ref{subsec:sed}). {\bf Lower}: average results from 2d fitting to the PACS and SPIRE maps of the extended emission (Sec.\,\ref{sec:dustmap}). }
\begin{tabular}{cccc}\hline
Epoch         &     M$_\mathrm{dust} $   & T$_\mathrm{dust}$    &    $\beta$     \\
                    &     M$_\odot$                  & K                              &                      \\ \hline
1                 &     0.54 $\pm$ 0.05        &  20.3 $\pm$ 0.7        &  2.2 $\pm$ 0.2       \\
2                 &     0.60 $\pm$ 0.05        &  18.3 $\pm$ 0.6       &  2.7 $\pm$ 0.2       \\
3                 &     0.60 $\pm$ 0.15        &  18.8 $\pm$ 1.9        &  2.6 $\pm$ 0.4       \\
all together & 0.57 $\pm$ 0.05        &  19.3 $\pm$ 0.7        &  2.43 $\pm$ 0.14       \\
\hline
1 & 0.57 & 20.7 & 2.22 \\
2 & 0.45 & 24.0 & 1.88 \\
3 & 0.54 & 23.5 & 2.12 \\ \hline
\end{tabular}
\end{center}
\end{table}

\paragraph{Warm point source}
Using only the \emph{Herschel} far-infrared observations of the central (unresolved) point source emission, we cannot extract values for the degenerate parameters of dust mass and dust temperature. Therefore we have extended the SED of the central point source to the near-infrared with 
{\it Spitzer} and Gemini mid-infrared results for the unresolved central source in 2007 (\citealt{2008ApJ...683L.171W}). We note that we exclude the {\it Spitzer} 2004/5 point source values which are several factors fainter for the emission from the central source (it is suggested this is because new dust was created in the point source in 2007; \citealt{2008ApJ...683L.171W}). The {\it Spitzer} 70\,$\mu$m point-source flux value from 2007 is 7.34\,Jy (\citealt{2008ApJ...683L.171W}) which is essentially the same as our measurements, giving some justification to combining the data of 2007 with our later epochs (2011/13).  Also available are  AKARI/IRC flux densities at 8 and 19\,$\mu$m (Ishihara et al. 2010), and while these wavelengths are expected to arise primarily from the central source it is not clear from the data how much extended source contamination could be present. 

We fit these data with a pure black body and with a ``dust'' emission model, that is using one or two modified black bodies similar to the procedure adopted for the extended emission. Since it was unclear which of the literature data we can safely use -- given it is unknown whether there were any intrinsic variations between then and now -- we chose to fit all possible combinations, and we report on the best of these here: the Gemini+{\it Herschel}+{\it Spitzer} data; the Gemini+{\it Spitzer} data; and {\it Herschel}+Gemini data.  The results of these fittings are given in Table~5 and Fig.\,6. 

\begin{figure*}[t!]
\begin{center}
\includegraphics[height=7cm]{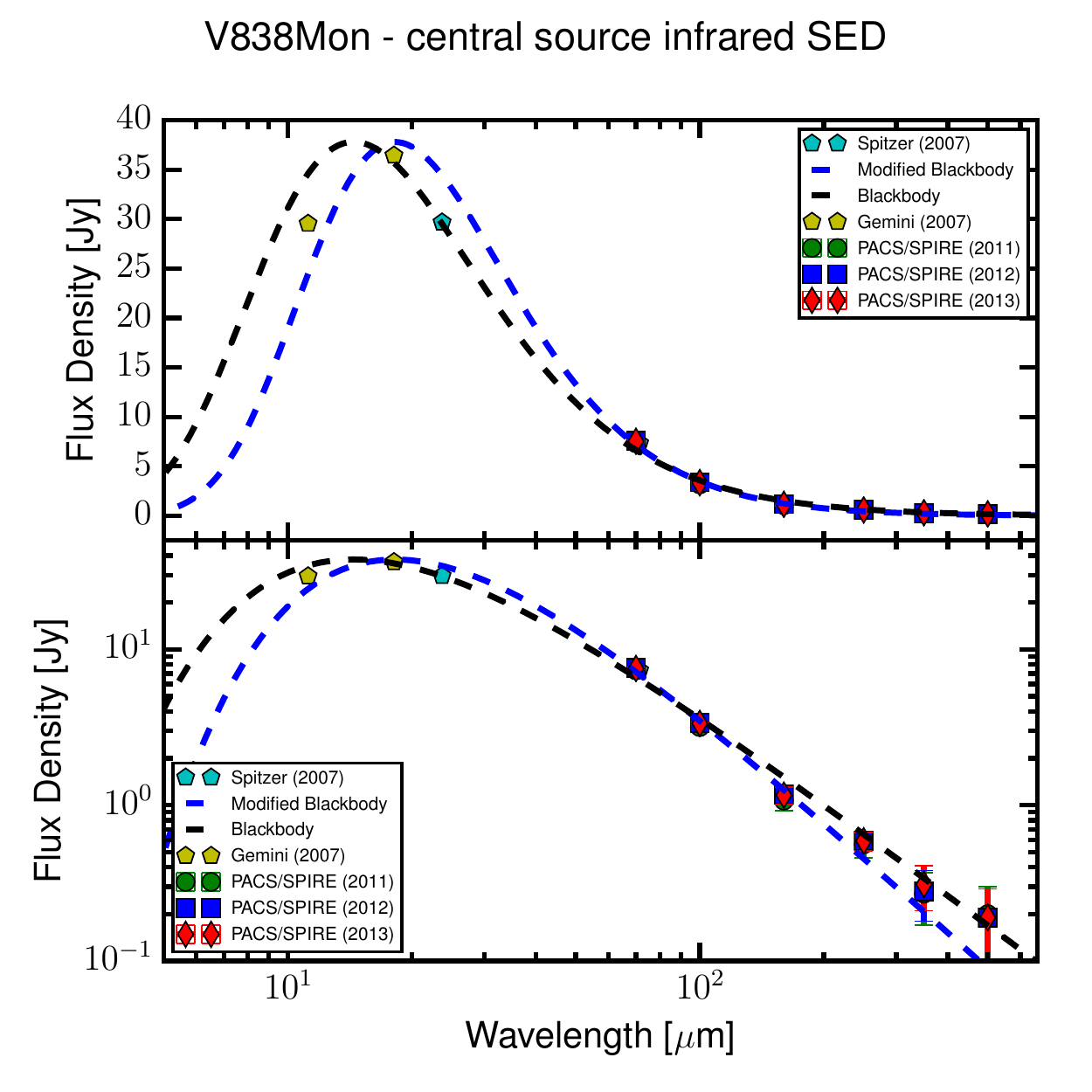}
\includegraphics[height=7cm]{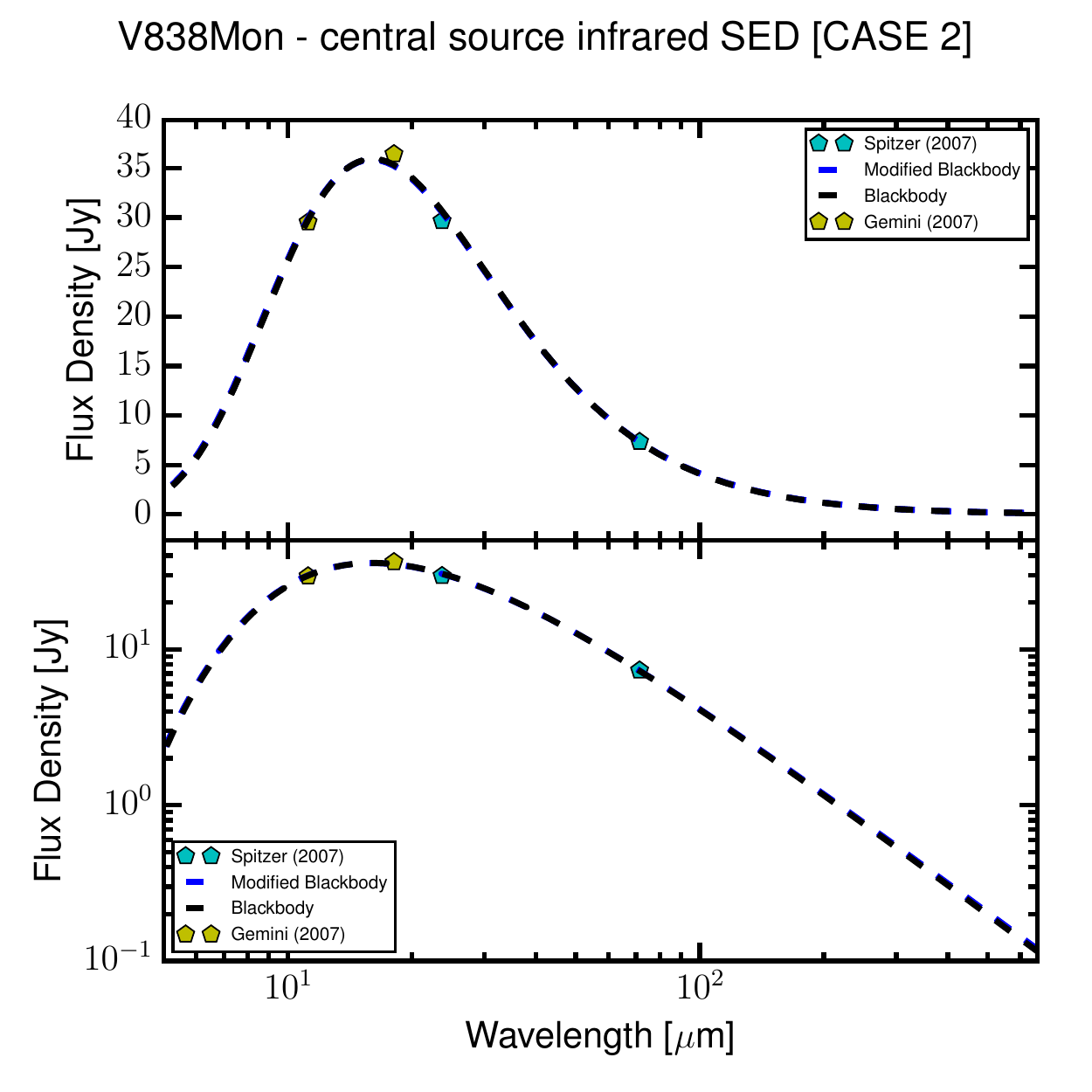}
\includegraphics[height=7cm]{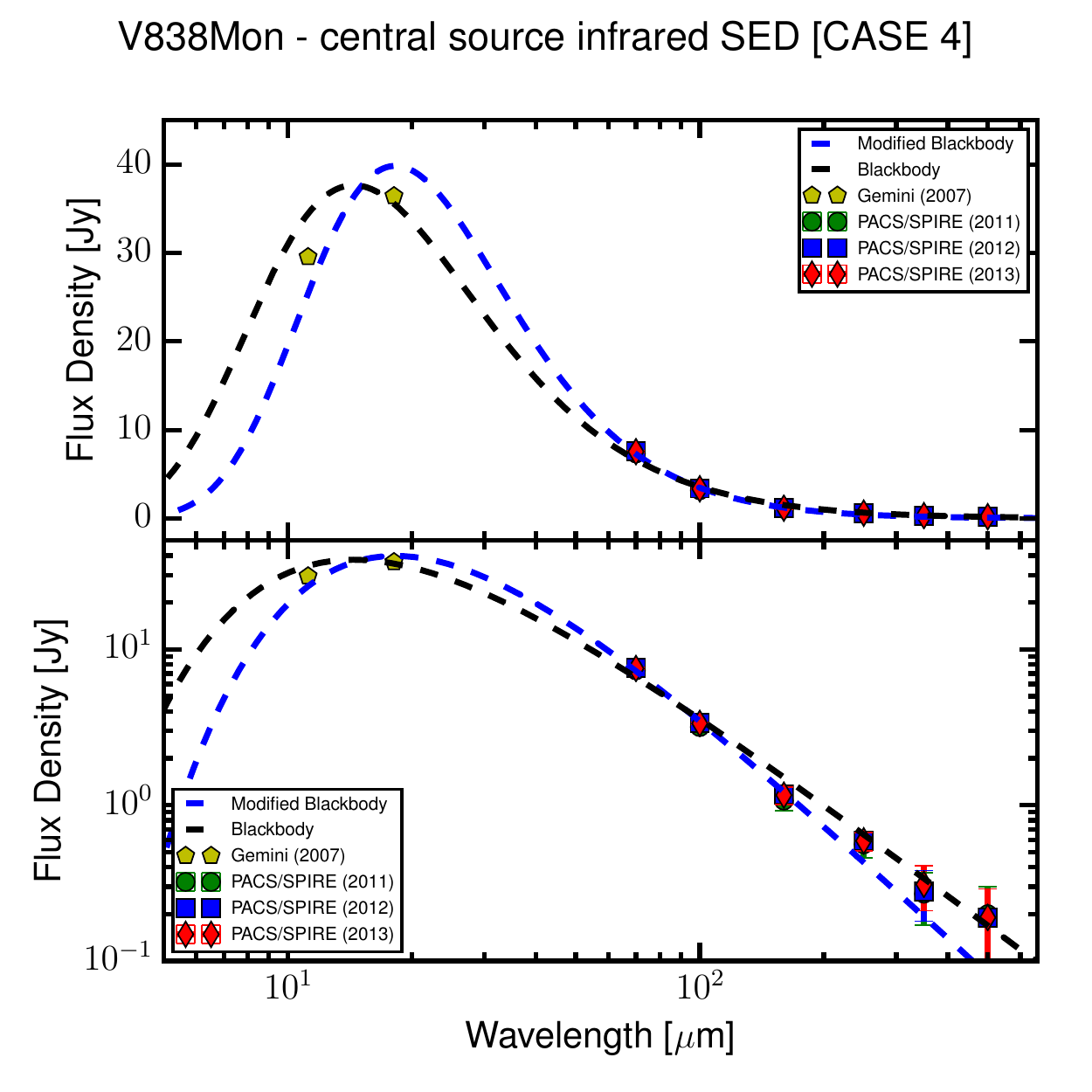}
\label{fig:sedherschel1}
\caption{Near- to far-infrared spectral energy distribution of the unresolved central point source, including flux densities measured with different combinations of {\it Spitzer}, Gemini, and {\it Herschel} data. } %2010A&A...514A...1I
\end{center}
\end{figure*}

The ranges of values presented in Table 6 show that the pure black body fits give a temperature in the range 300--350\,K and a radius of 150--200\,AU. The uncertainties on fit parameters are those provided by the least-square-fitting routine. The modified black body fit yielded less satisfactory results, so those of the simple blackbody are more secure, but with this modelling it is possible to estimate a mass of the material. The temperature range is a slightly lower, 230--330\,K, and the mass estimates lie around $9\times10^{-4}$\,M$_\odot$. The values for  $\beta$ are all not far from zero: that is close to a black-body. We note that the reliability of our mass estimate depends on the appropriateness of the dust model.

The uncertainties we quote in Table~5 are the statistical ones from the least-square fitting. The systematic uncertainties for this are much more difficult to ascertain and can be easily factor two to few, especially since we have no constraints from, for example,\ dust features.  A more elaborate model is beyond the scope of this work since we lack simultaneous multi-wavelength data at high spatial resolution and have in fact very limited information on the actual geometry and nature of this emission. 

\cite{Loebman:2015el} studied V838\,Mon with mid-IR data taken in 2008 and derive a blackbody temperature of the dust of 285\,K (since they have the shorter wavelengths their result is more secure than ours).

\begin{table}
\begin{center}
\label{tb:photextended}
\caption{Results of fitting models to the point source using the data from all Herschel epochs and data from the literature, and when using selected sets of data only}
\begin{tabular}{rcc}\hline
&Blackbody & Modified blackbody \\ \hline
\multicolumn{3}{c}{Gemini+{\it Herschel}+{\it Spitzer}}\\ \hline
M$_\mathrm{dust} $  (M$_\odot$)    & na & (8.3$\pm{0.6})\times10^{-4}$  \\
T$_\mathrm{dust}$ (K)   &353$\pm$31 &  237$\pm$24  \\
 $\beta$   &na &  0.43$\pm$0.12 \\
Radius (AU)& 153$\pm$9  & na \\ \hline
\multicolumn{3}{c}{Gemini+{\it Spitzer}}\\ \hline
M$_\mathrm{dust} $  (M$_\odot$)    & na & (1.1$\pm{0.1})\times10^{-3}$  \\
T$_\mathrm{dust}$ (K)   &317$\pm$6 &  321$\pm$24  \\
 $\beta$   &na &  -0.02$\pm$0.1 \\
Radius (AU)& 176$\pm$4  & na \\ \hline
\multicolumn{3}{c}{{\it Herschel}+Gemini}\\ \hline
M$_\mathrm{dust} $  (M$_\odot$)    & na & (7.9$\pm{0.5})\times10^{-4}$  \\
T$_\mathrm{dust}$ (K)   &353$\pm$36 &  234$\pm$49  \\
 $\beta$   &na &  0.49$\pm$0.14 \\
Radius (AU)& 153$\pm$11  & na \\ \hline
\end{tabular}
\end{center}
\end{table}

\subsection{Spatial information on dust properties}\label{sec:dustmap}

Dust temperature maps of the extended emission were created from pixel-by-pixel fitting of the measured flux densities with a modified black body ($S_\nu \propto B_\nu\,\nu^\beta$, as above). First the maps at 70, 100, and 160~$\mu$m were resampled (Swarp; \citealt{2002ASPC..281..228B}) to a common grid with identical pixel sizes, and subsequently convolved to match the point spread function (PSF) of either the 250~$\mu$m or 350~$\mu$m SPIRE photometry map. For this we adopted the convolution kernels and algorithm from \citet{2011PASP..123.1218A}. The sky background was computed in an ``empty'' region of the map and subtracted. A fit was then only attempted if the observed flux densities are higher than 10$\sigma_\mathrm{sky}$ of the sky background at at least three wavelengths.

We let $\beta$ run first as a free parameter and used the maps convolved to the 350\,$\mu$m SPIRE PSF. The resulting temperature, dust mass, and emissivity index ($\beta$) maps (epoch 1 only), as well as the corresponding percentage error maps, are shown in 
Fig.\,7.  
The noise-dominated low-emission edge regions suffer from degeneracy between dust temperature and dust emissivity, and so these regions have a higher uncertainty than the errors indicate.
In the next iteration we omitted the 350\,$\mu$m SPIRE data to improve the spatial fidelity of the dust images. To derive dust properties we use all maps convolved to 250\,$\mu$m. Again we limit the fitting to pixels with significant flux above the sky noise level at at least three wavelengths. In addition we fix $\beta$ to the value found in fitting also the lower resolution 350\,$\mu$m data. This approach improves the spatial resolution of the final dust maps, while retaining as much as possible information on the dust opacity (through $\beta$). Fixing $\beta$ to a single global value restricts the fits and limits the physical interpretation possible. This procedure was 
repeated for each of the three epochs.

The final dust temperature, dust mass and emissivity maps of V838\,Mon for each epoch are shown in 
Fig.\,8. 
Mild variations are seen between the three epochs, most noticeably in the temperature of epoch 2 (being slightly higher within the lobes) but the pixel value distribution of T$_\mathrm{dust}$, M$_\mathrm{dust}$, and $\beta$ agree with each other between the three epochs. 
The mean results from 2d fitting are given together with the ODR fitting results in 
Table\,4. 
The comparison in results for these two approaches is acceptable and we do not consider any of the differences to be significant.

\begin{figure*}[t!]
\begin{center}
\includegraphics[width=0.75\textwidth]{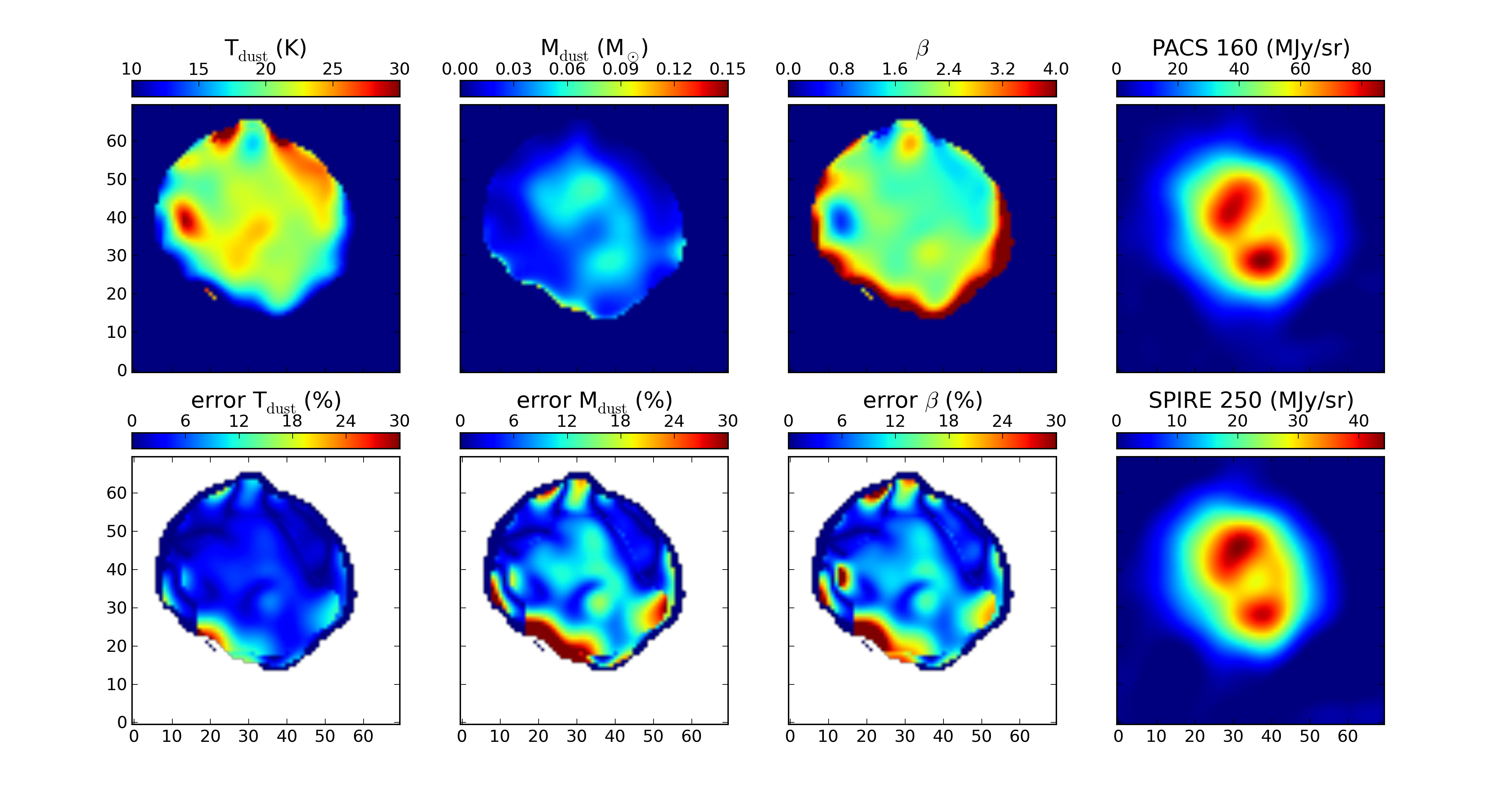}
\label{fig:dustmaps350}
\caption{{\bf Top panels}: Dust temperature, dust mass, and dust emissivity maps of V838\,Mon corresponding to the maps convolved to 350~$\mu$m and with $\beta$ a free parameter, for epoch 1. Colour-bars in each panel indicate the parameter range. Maps are 3.5\arcmin $\times$ 3.5\arcmin.
{\bf Bottom panels}: Corresponding percentage error maps. Orientation is the same as in Fig.\,1.}
\end{center}
\end{figure*}

\begin{figure*}[ht!]
\begin{center}
\includegraphics[width=0.75\textwidth]{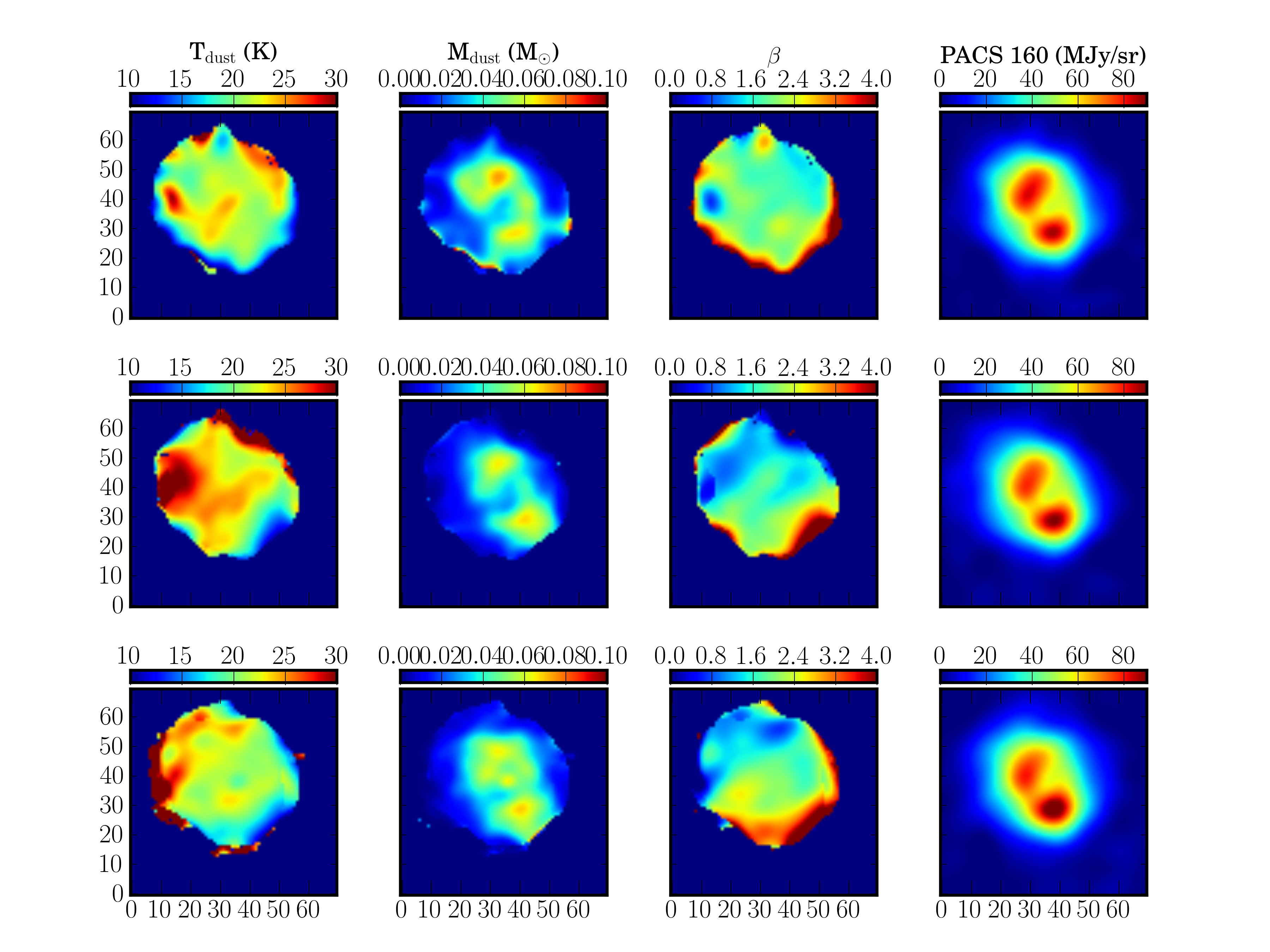}
\label{fig:dustmaps3epochs}
\caption{From left to right: Dust temperature (first column), dust mass
(second column), dust emissivity index (third column), and
PACS 160\,$\mu$m flux density (fourth column) maps of V838\,Mon created
from the maps convolved to 250\,$\mu$m and $\beta$ derived from the
350\,$\mu$m convolved data set.
From top to bottom: Epoch 1, 2, and 3. See text for details on the
procedure. Orientation is the same as in Fig.\,1.}
\end{center}
\end{figure*}

\subsection{Direct comparision with {\it Spitzer} maps}
\label{sec:spitzercomp}

The point source and extended source of V838\,Mon are very intimate at the FIR wavelengths. Even for the relatively small beam of {\it Herschel} it was necessary to use a point-source extraction method to separate the point-source from the extended source at the longer wavelengths; at the shorter wavelengths it was necessary to have the knowledge obtained from point-source extraction to find an appropriately small aperture with which to do aperture photometry. \\
{\sl For the point source:} With the 2005 {\it Spitzer} data, Banerjee et al.\ (2006) used a point-source extraction method to obtain $70$\,$\mu$m flux of $\sim4$\,Jy and a non-detection at 160\,$\mu$m ; with the 2007 {\it Spitzer} data, Wisniewski et al.\ (2008) used aperture photometry to obtain a $70$\,$\mu$m flux of $\sim7$\,Jy; and we measure $70$\,$\mu$m and 160\,$\mu$m fluxes of $\sim7$\,Jy and $\sim3$\,Jy.  \\
{\sl For the extended source:} Banerjee et al.\ report fluxes for the extended source at 70\,$\mu$m and 160\,$\mu$m of $\sim11$\,Jy and $\sim17$\,Jy, Wisniewski et al.\ make no specific statement (which implies a non-detection), and we measure $70$\,$\mu$m and 160\,$\mu$m fluxes of $\sim3$\,Jy and $\sim11$\,Jy.  The comparison of these epochs indicates that the point source flux increased with time and the extended source flux decreased.

In our opinion a comparison of the {\it Spitzer} and {\it Herschel} 160\,$\mu$m results should only be taken as indicative: the point source is so much fainter than the extended source and the {\it Spitzer} beam so large, that the reliability of the results is not high. But a comparison of the results at 70\,$\mu$m should be more reliable.

We decided to make a direct comparison of the total source (the point plus extended source) from the two epochs of {\it Spitzer} data with our {\it Herschel} data. The 2005 {\it Spitzer} maps that were used by Banerjee et al.\ are public. We downloaded these MIPS observations (AOR 10523648 and 10523904), and from the "maic" FITS files  we created single images at each wavelength using the Montage\footnote{http://montage.ipac.caltech.edu} code. We then convolved the {\sl Herschel} images to the beam size of the {\it Spitzer} data (18\arcsec\ at 70\,$\mu$m and 40\arcsec\ at 160\,$\mu$m) in HIPE, and converted to the MJy/sr units of the {\it Spitzer} data. 
A single value was subtracted from all maps to do a basic background subtraction. We then measured the fluxes from the {\sl entire} source in an aperture of 80\arcsec\ from each map (this being the aperture  used by Banerjee et al.).  We then did the same for the 2007 {\it Spitzer} data used by Wisniewski et al.\ (AORs  21476096, 21475840, and 21475072 from 2007, for 70\,$\mu$m and 160\,$\mu$m both).  Applying the appropriate colour corrections to the measurements is made difficult by the different temperatures of the point and extended sources (we measure here a {\sl total} flux), but in any case are only of the order a few percent. At 70\,$\mu$m we took a straight average of the respective corrections at 300\,K and 20\,K and at 160\,$\mu$m we used the cool value. 

We obtain: {\it Spitzer} (2005): 15.1\,Jy at 70\,$\mu$m, 16.1 at 160\,$\mu$m; {\it Spitzer} (2007): 12.4\,Jy at 70\,$\mu$m, 12.1 at 160\,$\mu$m; PACS (2011/12): 11.4\,Jy at 70\,$\mu$m, 15.4\,Jy at 160\,$\mu$m. 
(Banerjee et al.\ report fluxes for the 2005 data of 14.7\, and 17.5\,Jy; the slight difference with our measurements can easily be account for by our very rough background subtraction.)  We also compared all images directly to each other. The {\it Spitzer} images are slightly larger in the blue and slightly smaller in the red, but we conclude that there is no significant difference in the spatial extent of the emission between 2005 and 2011/12.

Previous investigations by the Herschel Science Centre showed that PACS and MIPS total flux and surface brightness agree within 5--20\%\ and no systematics were found\footnote{The calibration reports (PICC-NHSC-TR-034 and PICC-NHSC-TN-029) will be provided on the Herschel Legacy Library web-pages, which are currently being set up}. Taking into account also a 10\%\ calibration uncertainty quoted for both instruments, and our rough background subtraction, the differences in the total source flux that we measure (a drop with time at 70\,$\mu$m and drop then rise at 160\,$\mu$m) are barely significant. 

In conclusion: looking only at the 70\,$\mu$m results (where the point source and extended source are the least blended) the flux of the point source rose between 2005 and 2007 and then stayed the same, while the flux of the total source stayed the same (or dropped slightly) and its extent on the sky remained the same. Therefore, the brightness of the extended source must have dropped. Without longer wavelength multi-epoch data we cannot unfortunately say more.

\subsection{CO and SiO}\label{sec:cosio}

To interpret the line emission detected from CO,
we plot the energy diagram (Fig.\,9).  The data are taken from Table 3 and to the flux errors there we have added the calibration uncertainties and SPIRE continuum rectification uncertainty (Secs\,\ref{sec:pacsspec} and \ref{sec:spirespec}). 
The CO energy diagram shows two discrete components: a cold component with 
a peak of the line intensity at the upper state energy $E_{\rm up}<$55\,K (lower than J=4--3), and a warm component
with a rising trend with increasing $E_{\rm up}$.  Our working hypothesis is that the cold component is associated with ISM gas -- 
although it is also possible that it arises from a cold shell (\`{a} la \citealt{2004ApJ...607..460L}) -- while the warm component is associated with the circumstellar envelope/environment (i.e. arising out of the outflow triggered by the stellar impact). We stress that the spectra measured by PACS and SPIRE come from what they see as the point source, and at the spatial resolution at our wavelengths ($\sim9$\arcsec\ in at the shortest wavelength and $\sim37$\arcsec\  at the longest), this area is quite large (2.7 to 11\,pc at the distance of 6.2\,kpc).

We first compare the distribution of the CO fluxes on the energy-level diagram for V838\,Mon with those of what can be considered similar stars (i.e.\ having an extended atmosphere), and which are well-studied at these wavelengths: the red-supergiant, VY\,CMa and the AGB star, W Hya.  Here we are treating VY\,CMa and W\,Hya as templates  
of circumstellar envelopes created by a constant mass loss, and we wish to see how/where their mass loss differs to that of V838\,Mon. It is true that the mass loss from these stars has been ongoing for over a thousand years, a timescale much longer than that for V838\,Mon, but it is general trends we are looking for. It is also true that 
the mass-loss rate of VY\,CMa is highly variable (Decin et al., 2006; Matsuura et al., 2014). However, using the values in Decin et al.\footnote{Values we adopt for this back-of-the-envelope calculation are:  distance to the star= 1500\,pc, expansion velocity= 35\,km\,s$^{-1}$ (slower at the inner region), stellar radius=$1.6\times10^{14}$\,cm. The initial velocity acceleration region is at 10\,R$_\star$, with velocity  $\sim5$\,km\,s$^{-1}$, and the mass takes $\sim100$ years to cross. Then, from 10--100\,R$_\star$ we assume a constant mass-loss rate, with velocity gradually accelerating to $\sim$20\,km\,s$^{-1}$, and covering this distance takes 230 years. Finally,   
from 100--1000\,R$_\star$ there are several changes in the mass-loss rate, and this has a crossing timescale of $\sim1300$ years.} we find that only beyond 
$\sim1000$ \,R$_\star$ 
which takes a timescale of over 1000 years to reach, are the mass-loss rate variations likely to have an effect on the measurements. Hence we should avoid only this part of the comparison. 

The CO line intensities of these stars were taken from  \citet{Khouri:2014gz} and \citet{Matsuura:2013cz},
and these intensities have been scaled to the distance of V838\,Mon (6.2\,kpc); taking 1.14\,kpc for VY CMa (Choi et al.\ 2008) and 78\,pc for W Hya (Knapp et al.\ 2003, Justtanont et al.\ 2005). 
The current mass-loss rates of VY\,CMa and W Hya are estimated to be $2\times10^{-4}$\,$M_{\odot}$\,yr$^{-1}$,
and $2\times10^{-7}$\,$M_{\odot}$\,yr$^{-1}$.
The overall CO line intensities of the energy diagram reflect the mass-loss rates of these two objects.
The mass-loss rate of VY\,CMa is higher than W\,Hya, so VY\,CMa has brighter CO line intensities (after adjusting for its greater distance).
The energy diagram shows that V838\,Mon has fairly strong CO lines compared to these other two stars (scaled to the same distance).  

We can also see that V838\,Mon has a different CO curve to those of W\,Hya and VY\,CMa: it has a two-component curve, with a clear peak at $E_{\rm up}\sim50$\,K, and a gradual rise with increasing $E_{\rm up}$, while for the comparison stars there is an initial rise followed by a flattening. 
\citet{Khouri:2014gz} explained that higher $J$ (higher $E_{\rm up}$) transitions  
 trace the  inner and warm part of the circumstellar envelope, while lower $J$ transitions trace the outer and cooler part. The clear peak for V838\,Mon therefore lies part of the curve that traces the outer and cooler part, which would presumably correspond either to the first epoch of mass loss, that is,\ that occurring during the outburst, or to pre-existing material. The difference in the slopes at the higher $E_{\rm up}$ trace differences in the inner part of the circumstellar environment. The slow rise in the CO curve for V838\,Mon is because that the warm gas component is a single temperature zone, as our later fitting results will show: see also explanation Fig.\,10. For such a gas, the general driver of the curve on the CO diagram is the temperature, and it is the peak of the overall curve that indicates the gas temperature. The two evolved stars we compare to have a more complex story: for example, the temperature of gas from AGB stars is know to vary thanks to their more complex and longer  mass-loss history. Their curve on the CO diagram will therefore not be the same as that for a single-temperature zone.
 
 The curve of V838\,Mon shows that the molecular envelope consists of a cooler outer part and a warmer inner part, and the comparison to the evolved stars suggests that its envelope not have been made with a constant mass loss from the star. A good (and perhaps obvious) suggestion is that the material is associated with material erupted in 2002, which has remained close to the star and warm. The figure (specifically the comparison of the distance-scaled fluxes) also shows that the scale of gas eruption was large: one event of an eruption could have ejected material
equivalent to thousands years of constant mass loss from our comparison stars.

\begin{figure}[h!]
\begin{center}
\includegraphics[width=0.45\textwidth]{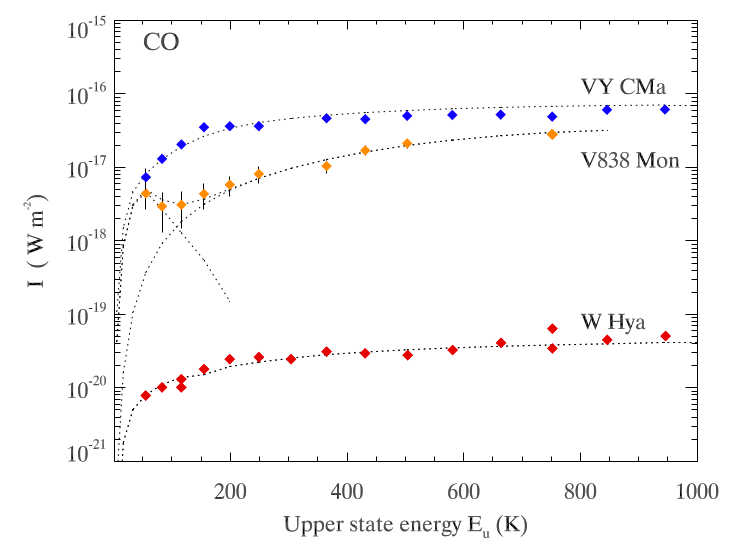}
\includegraphics[width=0.48\textwidth]{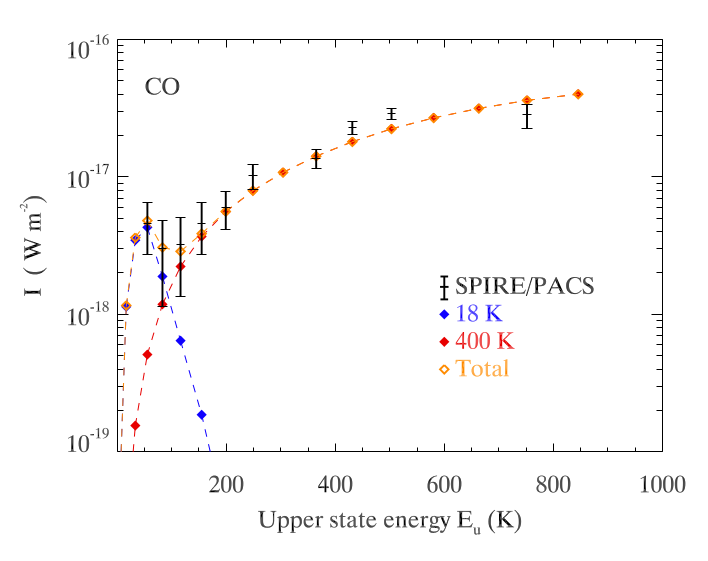}
\label{fig-co-comp}
\caption{{\bf Top:} The energy diagram for the V838 Mon CO lines, 
compared with the red superigant, VY CMa and the AGB star, W Hya.
The X-axis indicates the upper state energy of the CO transition ($E_{\rm up}$) in Kelvin,
and  the Y-axis shows the line intensities of the CO lines in W\,m$^{-2}$.
V838\,Mon shows two discrete components (warm and cold) of the CO energy distribution. Symbols are data and dotted lines are the fits (ours to V838\,Mon, and using the values from the respective papers for the other two stars). {\bf Bottom:} The energy diagram for CO lines detected by SPIRE and PACS (black symbols),
where  the X-axis shows the upper state energy of the CO transition ($E_{\rm up}$) in Kelvins,
and the Y-axis shows the line intensities of the CO lines in W\,m$^{-2}$.
Two discrete components of CO energy distributions
were detected, which were modelled with 18\,K (blue) and 400\,K (red).
The sum of these warm and cold components is plotted in orange.}
\end{center}
\end{figure}

We next derived physical conditions of the two CO-emitting components, by modelling the energy diagrams 
using the non-LTE radiative transfer code  {\sc RADEX} \citep{vanderTak:2007be}.
 {\sc RADEX} calculates the level populations of molecules, 
we used CO--H$_2$ and SiO--H$_2$ cross-sections and Einstein A-coefficients from the {\sc LAMDA} molecular and atomic data base of \citet{Schoier:2005ja},
which were adopted from calculations by \citet{Yang:2010bb} and \citet{Dayou:2006ey}.
We adopted a distance to the source as 6.2\,kpc.

\begin{figure}[h!]
\begin{center}
\includegraphics[width=0.4\textwidth]{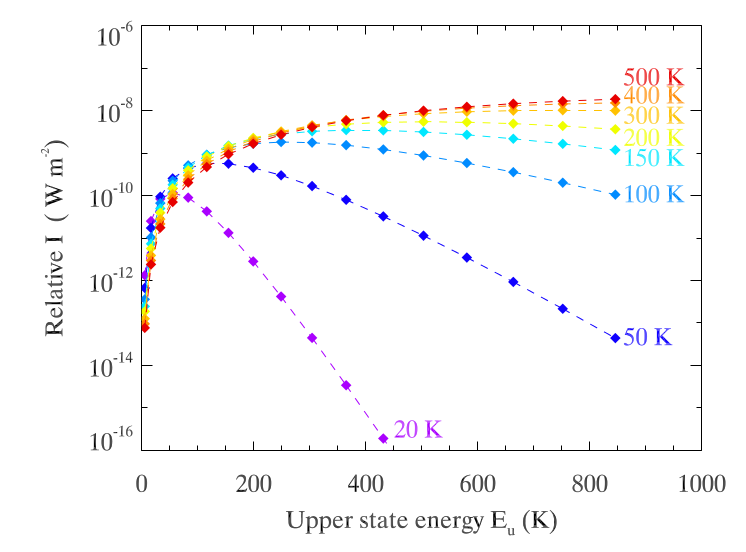}
\caption{Demonstration of how the CO energy diagram changes as a function of the kinetic temperature. This applies to the optically thin gas and with a constant beam size across all the  transitions, which is not the case for V838\,Mon. However, this figure is just to demonstration how the curve changes with temperature.}
\end{center}
\end{figure}

The parameters derived and the fitting errors are summarised in Table\,\ref{table-parameters}, and in Fig.\,9 we show the model fitted results. 
We now move our attention to a word about the fitting. The errors (1-$\sigma$ measurements+calibration error) are greatest for the cold component, and so simple $\chi^2$ fitting was weighted to the warm component and ignored the cold component. To overcome this limitation, we ($\chi^2$) fit the cold and warm components independently, and summed them. 
Strictly-speaking, a full radiative transfer modelling of the  cold component behind the warm component is  the correct path to follow, but given the gaps in our knowledge of V838\,Mon, we consider this an unnecessary step, and we ignore the small fraction of the ISM behind the warm component. The straight sum of the cold and warm components is therefore shown in Fig.\,9.  
Inspection of the fits on the plots resulted in some adjustment so we did not underestimate the cold--warm cross-over point at E$_{\rm up}\sim150$ K. The resulting $\chi^2$-fitting errors in the temperature are given in Table 6.  The errors for $N_{\rm CO}$ were calculated for a fixed temperature.

We derive temperatures of 400\,K and 18\,K for the two components. In this and the following discussion, we assume that these two components originate from different gas clouds: the cold component is ISM which is ubiquitous over the observed area (the point source diameter) and our results show it is optically thin, and the warm component comes from the stellar outflow and is smaller than the {\it Herschel} point source diameter.

\begin{table}
\begin{center}
\caption{CO model parameters with errors derived from the fitting. \label{table-parameters}}
\begin{tabular}{ccccc}\hline
  & warm & Cold \\ \hline
 $T_{\rm kin}$  (K)                      &  400 $\pm$ 50 & 18 $\pm$ 5 \\
 $\Delta \rm{v} (FWHM)$ (km\,s$^{-1}$)         & 200    & 5 (fixed) \\
 $N_{\rm CO}$  (cm$^{-2}$)     & $(1.0^{+4.0}_{-0.5})\times10^{20}$\tablefootmark{a}  & $(1.0^{+0.2}_{-0.5})\times10^{15}$ \\
 $M_{\rm CO}$  ($M_{\odot}$)  &$7.4\times10^{-5}$   &     &  \\ \hline
\end{tabular} 
\tablefoot{\\
\tablefoottext{a}{This range calculated for a temperature value fixed to that given in this table}} \\
\end{center}
\end{table}

\paragraph{Cold component}

The cold component could either be cold material associated with the outer parts of the molecular envelope, or it could be pre-existing ISM. Our data cannot distinguish between these two cases. 
The $\chi^2$ fitting gave a temperature of 18\,K, and  a best value for the H$_2$ density of $\sim1\times10^8$\,cm$^{-3}$ for the cold component, although this value is very poorly constrained and could be a factor of 100 lower. In the following we assume a constant density (which is a good assumption for ISM gas).  
The line width of the cold component has been fixed to 5\,km\,s$^{-1}$, as a typical
CO line width of molecular gas towards 
V838\,Mon was found to be 2--7\,\,km\,s$^{-1}$ from low $J$ CO lines \citep{Kaminski:2007gg}.
For the cold component, we assume that the SPIRE beams are filled by this cold CO gas,
and SPIRE beam sizes, which have a frequency dependence, are taken from \citet{Makiwa:2013ho}.
The estimated column density of the cold CO component is then $1\times10^{15}$\,cm$^{-2}$.

\paragraph{Warm component}

When fitting the warm component we find that CO line intensities are almost independent  of  the (assumed) H$_2$ density
when it is $>1\times10^6$\,cm$^{-3}$: anything between  $10^6$ and $10^7$\,cm$^{-3}$ is possible. The parameters resulting from the fitting are given in Table\,6. We note here that in the RADEX calculations, the escape probability used is that for a slab/cylinder morphology. \citet{2004ApJ...607..460L} suggests a model of a hollow sphere, and this model is one we wish to follow as much as possible in our calculations. 
The typical optical depth of the CO lines for V838\,Mon as estimated by the RADEX modelling of the Herschel spectra is 3. With this optical depth, the escape probability for a slab is about 30\% different to that for a sphere: this difference is very small, especially compared to our other uncertainties, and so we attempted no changes to the RADEX modelling process. 

With the fitting results, can we estimate the mass of the gas? This is a multi-step process, since we need to estimate the emitting area of the gas, which is not an output of the model fitting. 

First, can we estimate the FWHM of the CO emission lines? One of the model input parameters is  the line width (FWHM; $\Delta \rm{v}$), for which we did not have a direct measurement (but see below). The $\Delta \rm{v}$  of the warm component is linked with two key model parameters: 
 the CO emitting area, and  the saturation limit of CO lines.
 In the optically thin case, CO line intensities generally increase with higher column density.
Once the line becomes optically thick, the line intensities become saturated, and will not increase as the column density increases. 
Among commonly-observable molecules, CO lines tend to saturate with relatively small column densities.
The saturation limit is also coupled with the line width at a given column density.   
{\sc RADEX} defines the optical depth at the line centre ($\tau_0$) as
$\tau_0 \propto N / \Delta \rm{v}$, where $N$ is the column density
\citep{vanLangevelde:2012wu}.
Narrowing $\Delta \rm{v}$ results in lowering  the line intensities at the saturation limit.

A key point of  $\Delta \rm{v}$ is its link to the CO emitting area. The size of the gas shell created by the eruptive event in 2002  is limited by the expansion velocity and the time since the eruption. 
Let us assume that the warm gas has expanded at a constant velocity (v$_{\rm exp}$) since the eruption, and, as we will discuss later, v$_{\rm exp}=\Delta \rm{v}/2$. 
We will set the CO emitting area to be $\Omega_{\rm co}= \pi  \theta^2  / 4.25\times10^{10}$ sr,
where $\theta = \Delta \rm{v}/2 * t / d$ in arcsec, $t$ is the time since the eruption, $d$ is distance to V838\,Mon, and
$4.25\times10^{10}$ is the conversion factor between sterad and arcsec$^2$.
As the {\sc RADEX} output in erg\,cm$^{-2}$\,s$^{-1}$ is given as the emitting area for the {\sl whole} sky ($4\pi$),
to compare with the measured CO line intensities, which come from a smaller area than the whole sky, the model fluxes were  converted as 
$f_{\rm CO}$ (W\,m$^{-2}$)= $f_{\rm CO}$ (erg\,cm$^{-2}$\,s$^{-1}$) $\Omega_{\rm co} / (4\pi)$.  
Hence, the modelled values of $f_{\rm CO}$ (W\,m$^{-2}$) are $\propto \Delta \rm{v}^2$.
This analysis method follows the one used for the explosive event of
supernova 1987A (Matsuura et al. in preparation; \citealt{Kamenetzky:2013fv}).
This equation also gives the maximum possible line intensities 
for a given $\Delta \rm{v}$, because of the CO saturation limit.
Using this, we  narrowed down the range of acceptable $\Delta \rm{v}$ values as derived from our line fluxes.
By changing $\Delta \rm{v}$ over a 50\,km\,s$^{-1}$ grid,  
we established a limit to the $\Delta \rm{v}$ which translates to FWHM$>$200\,km\,s$^{-1}$: 
narrower than 200\,km\,s$^{-1}$ under-predicts the CO intensities. 

A value of  200\,km\,s$^{-1}$ is the lower limit for the intrinsic FWHM obtained from modelling  the fluxes, but what can we measure from the spectra? The lines in the SPIRE spectra, which has an instrumental resolution of 280--970\,km\,s$^{-1}$, are unresolved. One CO line is detected in PACS at 162\,$\mu$m  where the instrumental resolution is $\sim240$\,km\,s$^{-1}$. The line is rather noisy, but can be fitted with FWHM values of 240--325\,km\,s$^{-1}$ -- at this upper limit the intrinsic FWHM (decoupling the measured and instrumental widths) is $\sim220$\,km\,s$^{-1}$. So the upper limit line width measured from one CO line is close to the lower limit taken from the modelling of the fluxes. To be more certain whether the emission line widths support the modelling, it would be necessary to obtain higher resolution spectra.  For the calculations we do here, we will adopt a value of 200\,km\,s$^{-1}$ for the intrinsic  FWHM of the gas. 

Next, what expansion velocity does this FWHM translate into? The FWHM of a line arising from a hollow sphere (thin shell) is 2v$_{exp}$ (Robinson et al., 1982), and from a filled sphere is $\sqrt{2}$v$_{exp}$ (although the {\sl average} velocity could be lower; McCray, 1993). Taking a value of 200\,km\,s$^{-1}$ for the FWHM gives 
 an expansion velocity of the gas of 100\,km\,s$^{-1}$ for the hollow sphere case, and over 10 years this gives a radius of 210\,AU; for the filled sphere case the radius is 297\,AU. 
Expansion velocities of various values are reported in the literature as measured from different atomic and molecular lines, with values varying from 100--400\,km\,s$^{-1}$ (Loebman et al., 2015); Tylenda et al.\ (2009) measure a terminal velocity of 215\,km\,s$^{-1}$ in 2005. However, there are no contemporaneous measurements of the CO velocity for us to compare to.  

Finally, can we estimate the warm CO gas mass? We will estimate the CO mass as $r^2N_{\rm CO}\pi$, which equation assumes a slab geometry. \citet{2004ApJ...607..460L} suggests a model of a hollow sphere, however as long as the (averaged) density
along the line of sight remain the same, it does not matter if the geometry is a slab or a hollow sphere.  
With a column density of $1\times10^{20}$ and a radius of 210\,AU, 
 the CO gas mass is estimated to be $7.4\times10^{-5}$\,M$_{\odot}$ (see Table\,6).  We note however that the sources of uncertainty in this value are many -- the geometry, the radius, the uncertainty in the column density -- and this value should be taken as an estimate only.

\paragraph{SiO and H$_2$O}

A similar analysis  is applied to SiO, with {\sc RADEX} modelling (Fig.\,11) to all the measured transitions (measurement+calibration error included).
The collisional cross-sections and Einstein coefficients were taken from the {\sc LAMDA} molecular and atomic data base \citep{Schoier:2005ja},
which were adopted from calculations by  \citet{Dayou:2006ey}.
The SiO line intensities have  larger uncertainties than CO ones, 
and the transitions covered are $E_{\rm up}>380$\,K, much higher than CO.
Fig.\,11 shows that the kinetic temperatures of SiO could lie in the range 400--1200\,K,  and that the spectral
shape cannot be fit well with a simple component, but needs to involve multiple temperature
components.
The range of the temperature suggests that SiO has a a similar or higher temperature than the CO warm component -- we cannot constrain the options further, and can only state that the  SiO-emitting region is associated with CO-emitting regions or lies closer to the heating source, which is presumably the central star.

We did not carry out a similar analysis for the H$_2$O lines as they even more sensitive to the model used and assumptions adopted. 

\begin{figure}[h!]
\begin{center}
\includegraphics[width=0.45\textwidth]{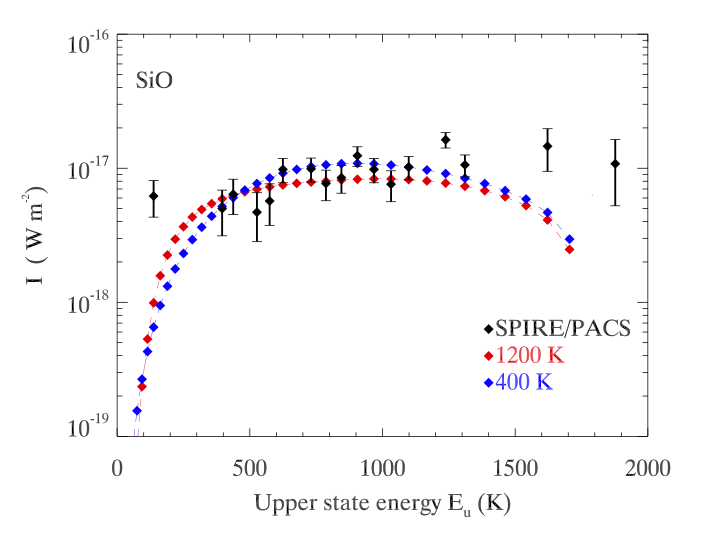}
\label{fig-sio}
\caption{The energy diagram of SiO (black symbols are data and measurement+calibration error) with {\sc RADEX} models with 400 and 1200\,K curves plotted.}
\end{center}
\end{figure}

\section{Discussion and conclusion}
\label{sec:discussion}

\subsection{Point source}

We remind the reader that the point source includes everything inside the PSF of PACS and SPIRE:
the PACS 70\,$\mu$m beam is  $\sim9$\arcsec, or 55000\,AU at the distance of V838\,Mon, and that of the SPIRE 500\,$\mu$m beam is  $\sim37$\arcsec. 

The SPIRE and PACS spectra show CO, SiO and H$_2$O molecules. 
The analysis of CO suggests presence of two component temperature, namely warm ($\sim$400 K) and cold ($\sim18$ K). 
The inferred SiO kinetic temperature is 400--1200\,K. 
In the past, molecules have been detected from V838\,Mon at optical, near- and mid-infrared and
millimeter wavelengths \cite[e.g.][]{2004ApJ...607..460L, Rushton:2005du}.
This includes the detection of CO rotational-vibration bands at near-infrared wavelengths \citep{2004ApJ...607..460L} 
and rotational transition at millimetre wavelengths  \citep{Kaminski:2007gg},
and rotational-vibration bands of SiO \citep{2004ApJ...607..460L}, as well as SiO masers \citep{2005PASJ...57L..25D}.
We will compare some of these past observations with our molecular line analysis.

\citet{2004ApJ...607..460L} used observations of V838\,Mon's IR spectra and proposed a model of a swollen star, with a cool and extended atmosphere and a circumstellar dust+gas shell; this model was expanded on using later observations by \citet{2007ASPC..363...39L} and then  \citet{Loebman:2015el}. According to the latter, the radius of the outer shell expanded, and the temperature dropped between 2003 and 2009: from 28\,AU to 263\,AU and 750\,K to 285\,K. \citet{Chesneau:2014jo} used interferometric measurements to determine that the star is surrounded by a compact, flattened dust structure, with the dust distributed from about 150 to 400 AU around the star (data from 2011--13).  They suggest that this structure is the dusty remains from the merger event that caused the outburst in 2002.

We estimated the radius of warm CO  to be 210\,AU, constrained by
PACS measurements of 162\,$\mu$m CO line width and CO saturation limit, and assuming the model of an thin spherical shell. Our estimated warm dust radius lies in the range 150 to 200\,AU, a value obtained by fitting our data with previously-taken literature data. These values are consistent with the findings quoted previously.

\citet{2004ApJ...607..460L} reported the detection of CO rotational-vibration bands
in the spectra up to 1.3 years after the outburst.
The estimated rotational-vibrational temperature of CO was 650\,K.
Our SPIRE and PACS spectra, which were obtained about nine years after the eruption,
showed the warm CO gas has a kinetic temperature of 400\,K.
These two measurements show that the temperature has dropped over the last eight years.  
If one assumes that the  eruptive shell is very thin, with a uniform density within, and that this shell has been expanding 
constantly since 2002, with about ten years between 2002 and our {\it Herschel} observations, the corresponding increase in volume and 
adiabatic cooling would normally lead to the temperature dropping by a factor of 100. That we do not see this suggests  that a nearby heating source is 
present.

V838\,Mon has a B3\,V companion, located 28--250\,AU distant (Munari et al., 2007; Tylenda et al., 2009). Could the companion be heating the gas? The obvious answer is "yes", given the luminosity of this type of star, however modelling its specific effect on the gas outflow from V838\,Mon is beyond the scope of this paper.  The gas temperature we measure is what we find, whatever the source of the heating. 
For the dust it {\sl is} possible to make a quick and simple calculation of the possible heating by the companion. Using equations from Tielens (2005: 5.43 and 5.44):
\begin{equation}
G_0 = 2.1\times10^4 (L_\ast/10^4L_\sun) (0.1\,{\rm pc}/d)^2 \\
\end{equation}
where we can adopt $L_\ast$=1000\,L\sun\ and $d=200$\,AU: the radiation field, G$_0\sim2.23\times10^7$. 
For amorphous carbon dust  ($\beta=1$) and taking a representative grain radius, a, of 0.1\,$\mu$m):
\begin{equation}
T_{dust} = 33.5 (1\mu\rm{m}/a)^{0.2} (G_0/10^4)^{0.2} \rightarrow T_{dust} \sim 248 \rm{K} 
,\end{equation}
or, for silicate dust ($\beta=2$): 
\begin{equation}
T_{dust} = 50 (1/a)^{0.06} (G_0/10^4)^{1/6} \rightarrow T_{dust} \sim 208 K 
.\end{equation}
Naturally the devil lies in the details. The grain radius, for example, depends on the environment in which they form --  but increasing the radius to 1\,$\mu$m only decreases the temperature by 1.6 and 1.15 for the two types of dust.  
These temperatures are close to the value we measure for the warm dust ($\sim300$\,K), and so we conclude that at least part of the source of the dust heating could be the companion star. 

One puzzle is why dust continued to cool from $\sim$750\,K to 280\,K in 10 years
as found by \citet{Loebman:2015el} and this work ($\sim300$\,K), while the gas 
temperature is slightly higher (400\,K).
One possibility is that  gas and dust do not have identical
temperatures, although this is more commonly found in low density regions with less frequent collisions, and is very unlikely to happen
in reasonably dense ($>1\times10^6$\,cm$^{-3}$) gas, such as V838\,Mon appears to have.

In our RADEX analysis we have adopted a spherical geometry. In fact this is probably not really the case. As well as the finding of \citet{Chesneau:2014jo}, evidence for a non-spherical morphology comes from \cite{2003ApJ...588..486W}. They found that spectroscopic and spectropolarimetric variations took place during the outburst, this being indicative of asymmetric geometry; this result is backed up the study of \cite{2004A&A...414..591D}, who also found polarimetric changes during 2002.  Further polarisation variations were observed a few months after the outburst by \cite{2003ApJ...598L..43W}, again suggesting the presence of an asymmetrical geometry of scattering material close to the star, and moreover that the distribution of this material had experience significant changes. It is therefore most likely that the circumstellar environment of V838\,Mon is not spherical. However, to conduct a more detailed modelling of the gas and dust in the region, its temperature and chemical structure, would require more information about this morphology.

We estimated a dust mass of $9\times10^{-4}$\,$M_{\odot}$ for the point source. This is based on a very simple modified blackbody fitting, and we note that the uncertainty in this value is high since it is obtained by fitting to previous epochs of data. \\
We estimated a CO mass of  $7.4\times10^{-5}$\,$M_{\odot}$, based on an assumed geometry and shell radius, and with factors of few uncertainty from the underlying CO column density. To turn this into a total mass we need a CO/H$_2$ ratio: we adopt here a ratio of $4\times10^{-4}$, which is the O-rich ratio for late-type stars (\citealt{Willacy:1997p27138}, and taken because of the physical rather than chemical similarity to V838\,Mon's extended environment). This then gives a total gas mass of  $\sim0.2$\,$M_{\odot}$. Taken together, these two values give a gas-to-dust mass ratio of 200. However, this value should only be taken as indicative, as the sources of uncertainty are large. If the outflowing gas from which the CO lines arise is molecular, the conversion from CO to total mass has only a factor of a few uncertainty. However, if the gas is neutral or ionised, then the total gas mass could be much higher

 If the outflowing gas from which the CO lines arise is molecular, our adopted value has only a factor of a few uncertainty. However, if the gas is neutral or ionised, then this value is completely wrong. Hence, a measurement of the gas-to-gust mass ratio requires more and better data. We also point out that it is not clear that the gas and dust we have detected arise from the same component: bear in mind the large size of the point source that {\sl Herschel} sees (the beam is 55,000\,AU at the shortest wavelengths). One should also consider that the dust created in the outflow from V838\,Mon -- created most likely as a consequence of an infalling low-mass star -- may not conform to the usual ISM. Little is known about the type and properties of dust created in these eruptive systems.

\subsection{The extended source}

The extended source detected in our PACS and SPIRE photometry is shown in the maps in Fig.\,\ref{fig:pext}. This region extends over  $\approx2.7$\,pc around V838\,Mon. The surface-integrated infrared flux (signifying the thermal light echo), and derived dust properties do not vary significantly between the different {\it  Herschel} epochs (1.5 years). Our multi-epoch photometry observations show that there has been no clear evolution in the morphology or the fluxes of either the point or the extended source over the 1.5 years. Our fluxes also do not differ strongly from what was found prior to our {\it Herschel} observations. \citet{Loebman:2015el} conclude that no new dust has formed around the point source since a  dust event taking place between 2004/5 and 2006/7 \citep{2008ApJ...683L.171W}, as they find no increase in the IR fluxes since then. Our 70$\,\mu$m fluxes are the same as the {\it Spitzer} 2007 value (\citealt{2008ApJ...683L.171W}), suggesting no further evolution has occurred.

We compared the fluxes of the emission of the total source (point+extended) between the PACS 2011/12 maps and the {\it Spitzer} maps of 2005 and 2007 (as used by \citealt{2006ApJ...644L..57B} and \citealt{2008ApJ...683L.171W}), and also compared the size of the emission from the extended source. There is slight evidence for a decrease in the 70\,$\mu$m flux from the total source between 2005 and then 2007/2011, but no difference in the extent of the emission when comparing the three epochs of maps at the same flux levels. However, \citet{2008ApJ...683L.171W} find  that the point source flux increased between 2005 and 2007 (from about 3.8\,Jy at 70\,$\mu$m to 7.3\,Jy) and our PACS 2011/12 value is the same as the {\it Spitzer} 2007 value. If the flux of the total source has not changed (or slightly dropped), and the extent of the emission from the extended source has not changed, this implies that the brightness of the extended source has dropped.

From the {\it Herschel} maps we  measured a dust mass of 0.57\,M$_\odot$, which while less than the 0.9\,M$_\odot$  measured by \citet{2006ApJ...644L..57B} (converted to a distance of 6.1\,kpc), is still high enough to result in a large gas+dust mass (if taking the standard ISM value of 100), and hence we agree with the conclusion that this material is ISM rather than arising from the star. 
We measured a temperature of $\sim19$~K dust for this source, which is only slightly warmer than the $T_{\rm kin}\sim12$\,K measured by Kaminski, Tylenda \&\ Deguchi  (2011) from CO lines from the ISM around V838\,Mon.

As mentioned in the introduction, \citet{2004A&A...427...193V} found evidence for multiple mass-loss events prior to the 2002 outburst: a 7\arcmin$\times10$\arcmin\ dust shell is visible in the IRAS data, and the MSX data also show an extended object (1.5\arcmin\ diameter) at the position of V838\,Mon. The latter exactly matches with the size of the infrared emitting region detected with \emph{Herschel}. The SPIRE images of V838\,Mon reveal patches of faint emission at $\sim$3\arcmin.5 south of V838\,Mon, but the maps are not large enough to verify if these structures are part of a larger structure corresponding to the IRAS dust shell.

With our PACS and SPIRE spectra we also detected a cold ($\sim18$\,K) component of CO gas, which has a temperature close to that measured for the dust. 
This low temperature suggests that the cold component is associated with ambient ISM gas. 
The ISM is also responsible for millimetre CO lines which were detected by \citet{Kaminski:2007gg}, 
and are most likely associated with the light echo detected by
\citet{2003Natur.422..405B}. 
 \citet{2011A&A...529A..48K} found the kinetic temperature of CO to be 12--15$\pm5$\,K, and 
our kinetic temperature of CO gas is slightly higher but consistent within the uncertainty.
There is a large discrepancy in the column density of the cold CO component: 
in the range 2--6$\times10^{16}$\,cm$^{-2}$ from  \citet{2011A&A...529A..48K} 
while we obtained $1\times10^{15}$\,cm$^{-2}$.  
Some of the discrepancy could be associated with the uncertainty
of H$_2$ density, as a higher density of the collisional partners would require a lower column density
to reproduce the same line intensity within that range (this does not apply to warm component). Another possibility is that the cold CO is not from the ISM, but rather from a cold shell that is associated with the warm shell (e.g.\ the \citealt{2004ApJ...607..460L} model), and hence is part of the {\it Herschel} point source.

Our {\em Herschel} study of V838 Mon has left us with many questions. What is the relationship between the gas and the dust -- are they co-spatial, and how were they each created by the outburst or within the outflow? How many temperature zones are present in the outflow, and does the cold CO gas arise from the ISM or is it directly related to the warm gas? How much dust has been created in the outburst outflow? When will the star return to its pre-outburst state? All very interesting questions for the future.

\begin{acknowledgements}
We thank John Wisniewski for a thoughtful refereeing process that resulted in an improved paper.
NLJC and KME acknowledge support from the Belgian Federal Science Policy Office via the PRODEX Programme of ESA.
PACS has been developed by a consortium of institutes led by MPE (Germany) and including UVIE (Austria); KU Leuven, CSL, IMEC (Belgium); CEA, LAM (France); MPIA (Germany); INAF-IFSI/OAA/OAP/OAT, LENS, SISSA (Italy); IAC (Spain). This development has been supported by the funding agencies BMVIT (Austria), ESA-PRODEX (Belgium), CEA/CNES (France), DLR (Germany), ASI/INAF (Italy), and CICYT/MCYT (Spain).
SPIRE has been developed by a consortium of institutes led by Cardiff University (UK) and including Univ. Lethbridge (Canada); NAOC (China); CEA, LAM (France); IFSI, Univ. Padua (Italy); IAC (Spain); Stockholm Observatory (Sweden); Imperial College London, RAL, UCL-MSSL, UKATC, Univ. Sussex (UK); and Caltech, JPL, NHSC, Univ. Colorado (USA). This development has been supported by national funding agencies: CSA (Canada); NAOC (China); CEA, CNES, CNRS (France); ASI (Italy); MCINN (Spain); SNSB (Sweden); STFC and UKSA (UK); and NASA (USA).
This research has made use of the SIMBAD database, operated at CDS, Strasbourg, France.
Herschel is an ESA space observatory with science instruments
provided by European-led Principal Investigator consortia
and with important participation from NASA.

\end{acknowledgements}

\bibliographystyle{aa}  % aa.bst
\bibliography{v838mon3_v5.bbl} % 

\appendix

\section{Footprint of the PACS and SPIRE spectrometers}\label{sec:app3}

\begin{figure}
  \centering
  \includegraphics[scale=0.25]{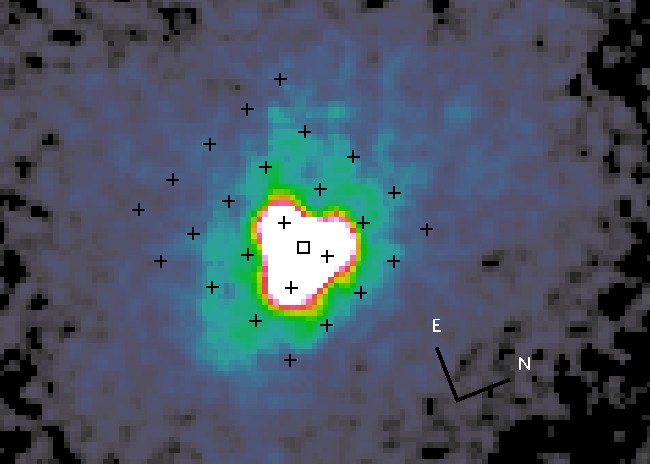}
    \includegraphics[scale=0.25]{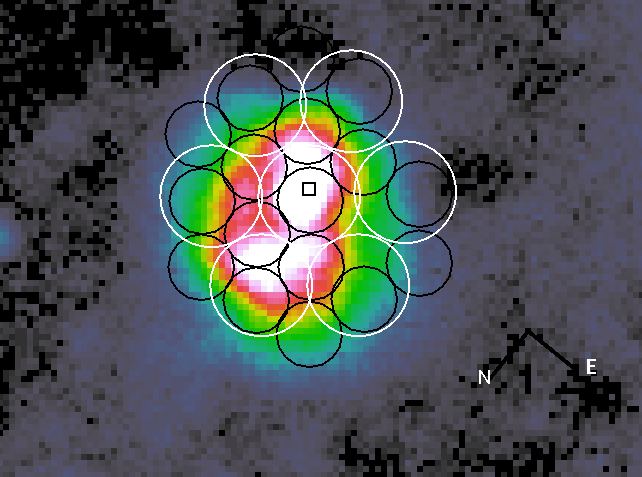}
   \caption{{\bf Top}: crosses mark the centre of each of the PACS spaxels on the 160\,$\mu$m map. {\bf Bottom} the non-vignetted circles of the SSW bolometers (black) and the SLW bolometers (white) on the 250\,$\mu$m map. The position of V838\,Mon is marked with a square.}
\label{fig:footprint}
\end{figure}

As mentioned in Secs\,\ref{sec:pacsspec} and \ref{sec:spirespec}, the footprint of the spectrometers extends over the extended emission around V838\,Mon, and some of this extended emission ``contaminates'' the point source. In Fig.\,\ref{fig:footprint} we show the footprint of the PACS and SPIRE spectrometers plotted on two of our epoch 1 maps. 

\section{Producing a calibrated point-source PACS spectrum}\label{sec:app6}

The HIPE task {\it extractCentralSpectrum} works by adding together the spectra of the spaxels that contain the flux of the point source, and then applying a correction for the shape of the beam (which is much wider than a single spaxel): effectively it performs a wavelength-dependent aperture correction to the flux levels in the spectrum. This correction, contained in the PACS calibration tree, is to be used on sources located close to the centre of the central spaxel of the 5x5 IFU, in particular because only for this central spaxel has the PACS beam been well calibrated (private communication). The task can extract only the central spectrum or the sum of the central 3x3 spaxel box, and the point-source flux density loss correction applied takes this into account. For our data of V838\,Mon, the star was located almost exactly between the central spaxel and a neighbour, and hence the task provided could not be used. However, the same {\sl procedure} could be used, and the resulting spectrum would then be more correct than a simple combination of the brightest three spaxels which contain the flux of V838\,Mon (private communication with the PACS instrument team).  

Therefore we modified the procedure  of {\sl extractCentralSpectrum} to take into account our non-central position for V838\,Mon. For each of our 20 spectral segments we did the following: 
\begin{enumerate}
\item Extracted the summed spectrum of the 3x3 spaxel box surrounding our brightest spaxels (those with most of the flux from V838\,Mon).
\item Multiplied that by the values in the point-source flux-loss correction tables from the calibration tree ({\sl calTree.spectrometer.pointSourceLoss}  using the central-3x3-to-total values).
\item Then extracted the spectrum only of the brightest two spaxels within the previous 3x3 spaxel box: this spectrum has a measurably superior signal-to-noise ratio than that of step 1.
\item Fit the continuum of the spectra from steps 2 and 3, and divided the first by the second.
\item Then multiplied that ratio into the spectrum from step 3. 
\end{enumerate}
The result is a spectrum with the best possible SNR but corrected for the point-source flux density losses.

\section{Using deconvolution to assist with the point source photometry}\label{sec:decon}

The main difficulty with measuring the point source and extended source fluxes separately from the PACS and SPIRE maps is that the point source is located within the extended emission. For the PACS maps it is possible to still see the two separately, but for the SPIRE maps -- with a fainter point source and a larger beam -- the two are particularly intimate. 

To try to improve the results of the PSF subtraction, we tested our PSF extraction method (App.\,\ref{sec:app5}) on maps which had first been deconvolved: the separation between the point and extended source will then be greater and PSF subtraction should be easier. As a bonus, we would also have deconvolved images of the extended IR bright region to study.

The deconvolution was based on a maximum entropy method (MEM) and follows in general the scheme described by Hollis et al.\ (1992) for HST images and Ottensamer et al.\ (2011) for Herschel/PACS. In the deconvolution task, the multiplier is convolved with the PSF model. We tested with the PACS and SPIRE beams (as used also in our PSF-subtraction work) but chose instead to use observations of the point source AFGL\,3068, which shows no circumstellar emission and gave very clean results. In the beginning the multiplier is the PSF itself, which is then scaled by the flux ratio of the image and the PSF, and finally re-convolved. At each iteration step the fluxes are compared and the residuals become part of the new multipliers. The iteration is stopped when artefacts become visible, which are mainly manifest as negative fluxes around the central source.  The deconvolution was applied to the PACS and SPIRE maps of V838\,Mon created by Scanamorphos, and we also deconvolved the maps of AFGL\,3068 to then use them for the subsequent PSF-subtraction photometry, using the same technique as outlined in App.\,\ref{sec:app5}.

The photometry resulting from the PSF subtraction method on the deconvolved maps is reported in Table\,\ref{tab:deconvolvephot}. 
Unfortunately, this process did not offer as much of an  improvement as we had hoped. For the PACS 70\,$\mu$m and 100\,$\mu$m maps deconvolution did not make a much difference, and the problem of residuals resulting from a mis-match between the adopted PSF shape and the real PSF for our  V838\,Mon observations remained (see App..\,\ref{sec:app5}). For the PACS 160\,$\mu$m maps the deconvolution resulted in a slight improvement: the scatter in the resulting photometry is lower, and the results agree better with the aperture photometry (which are the preferred results for PACS) reported in Table\,\ref{tab:pphot}.  For the SPIRE maps, doing PSF subtraction on the deconvolved maps was a challenge,  partly because the deconvolution appeared to alter the FWHM of the PSF differently to that of the V838\,Mon maps -- an extra smoothing step was necessary for the subsequent PSF subtraction. The photometry obtained from the PSF subtraction on these maps is in the same range as reported in Table\,\ref{tab:pphot}, however, the 500\,$\mu$m values are lower than those found previously, and this probably does indicate that our previous results are slightly too high.

In summary, using deconvolution to aid in the PSF subtraction did not help as much as hoped in our work because it was clear that a {\sl very good} knowledge of the beam for your particular observation (i.e. subjected to the same observing plan, the same data reduction, and the same type of map-making) is crucial to achieving both a good deconvolution and a good PSF subtraction. Given the aims of this paper, we did not feel it was justified to spend more time on the deconvolution. However, in our opinion this method would work if more time could be invested in it.  

\begin{table}[ht!]
\caption{PACS and SPIRE point source fluxes as obtained from PSF subtraction on deconvolved maps. Beam and colour corrections have been applied.}
\label{tab:deconvolvephot}
\centering
\begin{tabular}{lc}\hline
Band ($\mu$m) & (Epoch) Flux density (Jy) \\ \hline
70 & (1) 6.7--7.3 (2) 7.0--7.6 (3) 7.0--7.6 \\   
100 & (1) 3.8--4.1 (2) 3.8--4.1 (3) 3.6--3.9\\      
160 & (1) 1.2--1.3 (2) 1.3--1.4 (3) 1.2--1.3\\     
250 & (1--3) 0.53--0.62  \\       
350 & (1--3) 0.21--0.27\\                  
500 &  (1--3) 0.08-0.15 \\          
\hline
\end{tabular}
\end{table}

\begin{figure*}
  \centering
      \includegraphics[scale=0.5]{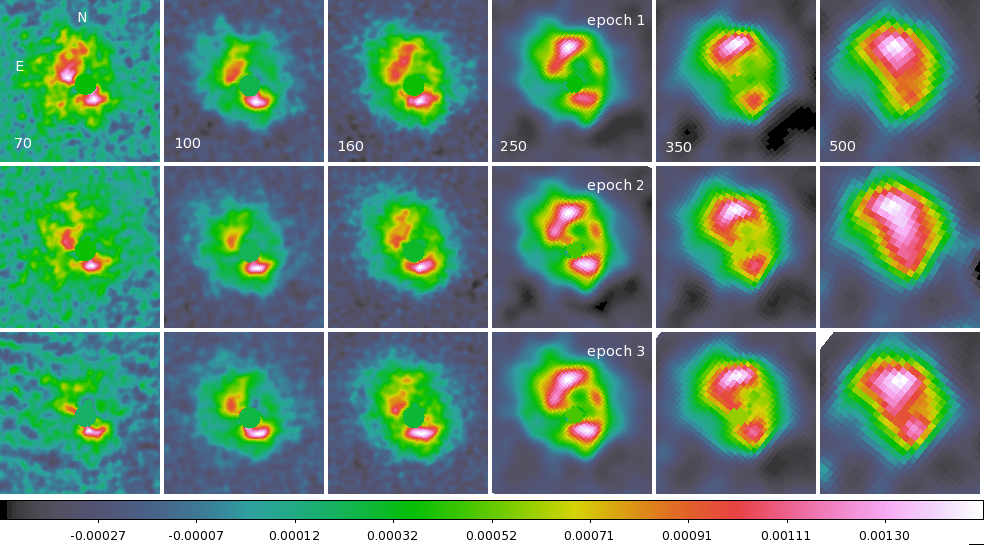}
   \caption{The deconvolved maps with the point source subtracted and with the subtracted region replaced with a constant value. Epochs, wavelengths, and orientation are indicated on the maps. Scaling is min--max for each image.}
   \label{fig:deconvolve}
\end{figure*}

The images resulting from the deconvolution are presented in Fig.\,\ref{fig:deconvolve}. For these we have subtracted the point source.  It is clear that the overall features at each wavelength do not change with epoch, with the possible exception for the PACS maps, where it appears that the northern lobe becomes fainter with time. 

\section{Photometry corrections}\label{sec:app1}

For aperture photometry of the point source for the PACS and SPIRE data we used the aperture sizes and corrections given in Table\,\ref{tab:photcorrs}. The  PACS aperture corrections were taken from {\it photometer.apertureCorrection}  in calibration tree 65 (the fm7 values), except for the aperture photometry for the scaled PSF maps, for which we used  apertures of 57.6\arcsec, and the appropriate aperture corrections provided in the ASCII files that accompanied the tarball of the beam files (see below for details of the beams). The colour corrections for PACS were taken from report on the PACS calibration wiki page (dated April 12, 2011: cc\_report\_v1.pdf), and the colour and beam-area corrections for SPIRE were taken from the SPIRE calibration tree and a SPIRE photometry script provided in HIPE.

\begin{table*}
\begin{center}
\label{tab:photcorrs}
\caption{Details of the apertures and the multiplicative corrections used in the photometric measurements of the point source}
\begin{tabular}{cccccc}\hline
band & ap.\ radius & ap.\ correction  & \multicolumn{2}{c}{colour correction\tablefootmark{b}} & beam-area correction\tablefootmark{a} \\ \cline{4-6}
& & & point  & extended  &\\ \hline
PACS 70&6.06\arcsec    & 1/0.639 &  1/1.005 &  1/1.224&na \\
PACS 100&4.95\arcsec  &  1/0.516 & 1/1.023& 1/1.036&na\\
PACS 160&4.03\arcsec & 1/0.211   &1/1.062&1/0.963&na\\
SPIRE 250 & 22\arcsec & 1.2614& 0.945& 0.945&1.0451 \\
SPIRE 350 & 32\arcsec &1.2258& 0.948& 0.948&1.0433\\
SPIRE 500 & 42\arcsec &1.2015& 0.943& 0.943&1.0756\\
\hline 
\end{tabular}
\tablefoot{\\
\tablefoottext{a}{to correct to a point source with a spectral index of $\alpha=2$, rather than -1, as is applied in the pipeline}\\
\tablefoottext{b}{assuming a spectral index of $\alpha=2$ for  SPIRE; assuming a PACS point source of 250\,K and extended source of 20\,K} \\
}
\end{center}
\end{table*}

\section{Details on the PSF-subtraction method}
\label{sec:app5}

For performing PSF subtraction using the scaled PACS/SPIRE beams, we took the beams from:
\begin{itemize}
\item PACS: a tarball of FITS files (``PACSPSF\_PICC-ME-TN-033\_v2.0'') from the PACS documentation web-site from the \emph{Herschel}  portal.  We used the examples created from Vesta, with  scan speed 20, pixfrac 0.1, from OD 345, and with array-to-map angle +42.
\item SPIRE:  the normalised beams created in 2012 and provided on the SPIRE documentation web-page from the \emph{Herschel}  portal (specifically the page: herschel.esac.esa.int/twiki/bin/view/Public/ SpirePhotometerBeamProfile. 
\end{itemize}

The steps were:
\begin{enumerate}
\item Resample the PSF maps to the pixel sizes of the V838\,Mon maps.
\item Rotate the PSF maps to match the position angle of the V838\,Mon maps: the shape of the PSF depends on the scanning speed, angle, and the position angle (this is more true for PACS than for SPIRE). We note that rotation was done on the WCS, rather than on the map as an image.
\item Measure the sky position of the PSF and of V838\,Mon to within a pixel, using {\it sourceExtractor}.  Change the WCS values in the PSF maps so the PSF star and the point source of V838\,Mon are in the same position.
\item Scale and subtract the fluxes of the PSF maps from the V838\,Mon maps (in WCS space, using the HIPE task {\sl imageSubtract}).
\item Determine the best scale-and-subtract value via visual inspection of the residual maps and plots of cuts taken at various position angles through the residual maps. The acceptable scaling factors are those that result in residuals that are the flattest.
\item Finally, replace the fluxes in a small circle around the subtracted star with the value of the flux surrounding that region, to create cleaner PSF-subtracted maps. These were used to create residual-free images for this paper, and to use as a check on the extended-source photometric measurements (Sec.\,\ref{sec:photmease}).
\end{enumerate} 

\begin{figure*}
  \centering
      \includegraphics[scale=0.25]{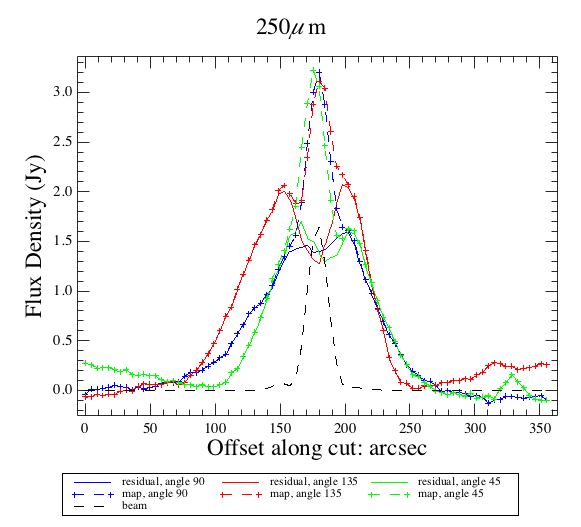}
      \includegraphics[scale=0.25]{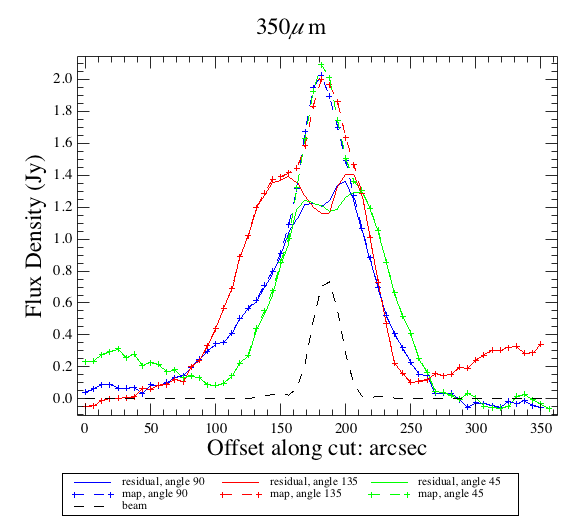}
      \includegraphics[scale=0.25]{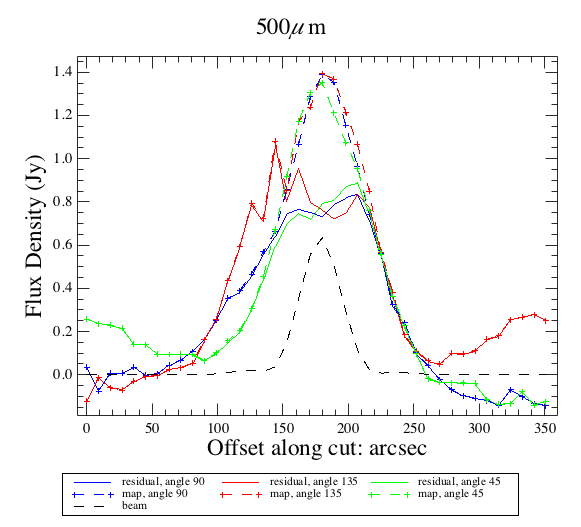}
       \includegraphics[scale=0.25]{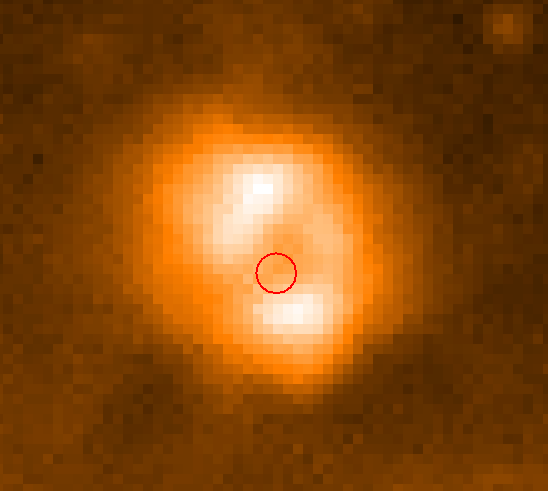}
       \includegraphics[scale=0.25]{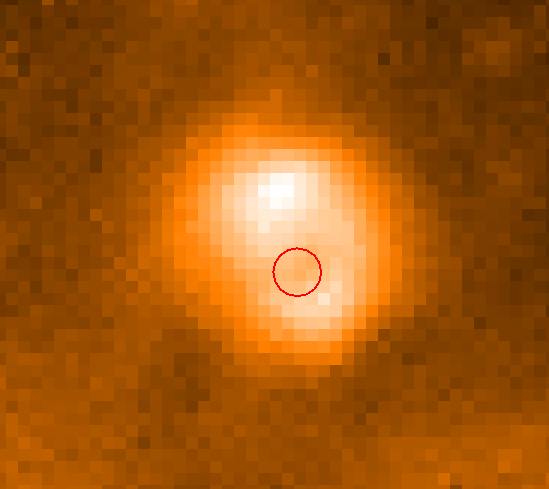}
       \includegraphics[scale=0.25]{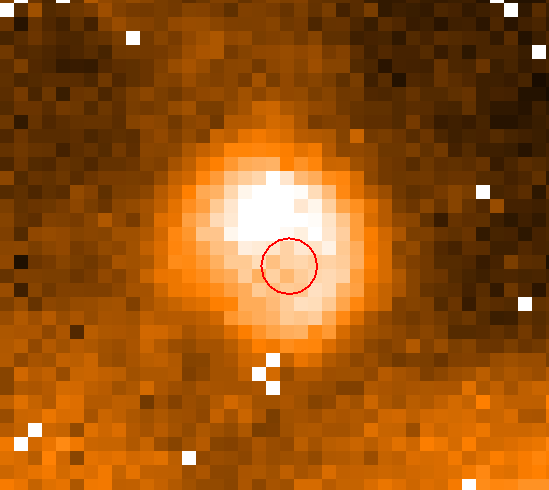}
   \caption{Examples of subtracting the scaled PSF from the epoch 1 SPIRE maps. {\bf Top} are cuts through the original (coloured crosses) and the residual (coloured dashed) maps at various position angles (blue: $90^{\circ}$, red: $135^{\circ}$, green: $45^{\circ}$) and a cut though the scaled PSF at position angle 0 (black dashed). {\bf Bottom} are the residual maps with the position of V838\,Mon marked.}
   \label{fig:resid}
\end{figure*}

Fig.\,\ref{fig:resid} shows the SPIRE maps after having scaled-and-subtracted the PSF (Sec.\,\ref{sec:spirephot}). Also shown are cuts taken through the original and the residual maps at various position angles, and a cut through the scaled beam. It is from these plots and figures that the decision about the best scaling factors was made. 

The PSF-subtraction process did not work for PACS as well as it did for SPIRE. We always had negative residuals, arising from a difference in the shape of the beam from the PSF maps and the effective shape of beam for our V838\,Mon maps. It is noted in the report accompanying the PACS beams that the beam-shape depends also on the data reduction and map-making so that a re-reduction of the PSF data, following the same reduction and map-making methods used for the astronomical observations, is recommended. However, as we use the PSF-subtraction method for the PACS data only as a check, we did not do this.  
An additional consideration is that it was necessary to rotate the beams (the supports of the {\it Herschel} secondary lead to a tri-lobal shape for the beams, which orientation depended on the position angle). Rotation inevitably involves an interpolation of the images, and it is possible that this could also account for the mismatch between our point source and the beam profiles. The tri-lobal shape (and hence the resampling of that pattern) is more prominent in the PACS than the SPIRE beams, and this could also explain why the mismatch is greater for the PACS beams.

\section{PACS spectral lines without a clear identification}\label{sec:appunid}
Lines from the PACS spectra for which we could not find a clear identification, but for which the detection is formally significant, are given in Table\,\ref{tab:unid}. We checked these all against a basic model spectrum based on the parameters of our point source, to look for more CO and H$_2$O lines, but no matches could be found. Other potential lines include SiO, CO+, and SO$_2$, but a proper modelling of these lines is necessary before this can be taken further. 

\begin{table}[ht!]
\caption{PACS spectral lines without a clear identification. }
\label{tab:unid}
\begin{tabular}{lclc}\hline
Wavelength      &       Flux& Wavelength        &       Flux  \\
($\mu$m)&(10$^{-17}$ W/m$^2$) & ($\mu$m)&(10$^{-17}$ W/m$^2$)   \\\hline
70.74   &  $5.26\pm{2.59}$ &159.12    &    $1.87\pm{0.18}$   \\
71.54       &  $3.97\pm{0.52}$    &159.32   &     $1.98\pm{0.18}$  \\
71.56      &   $6.99\pm{0.51}$    &159.47     &   $3.06\pm{0.20}$   \\
81.61\tablefootmark{a}    &$1.66\pm{0.71}$   &162.37    &    $2.13\pm{0.45}$   \\
81.73     &    $3.45\pm{0.61} $  &165.11   &     $0.80\pm{0.10}$   \\
83.00\tablefootmark{a}   & $11.54\pm{1.00}$  &179.96 &      $1.10\pm{0.09} $  \\
84.80   &$10.38\pm{1.40}$    &181.12 &   $1.05\pm{0.11} $  \\
90.03   &   $1.47\pm{1.90}$ & & \\ \hline
\end{tabular}
\tablefoot{
\tablefoottext{a}{wide line}
}
\end{table}

\label{lastpage}
\end{document}